\documentclass[%
 reprint,
superscriptaddress,
nofootinbib,
nobibnotes,
 amsmath,amssymb,
 aps,
 prd,
]{revtex4-1}

\usepackage{natbib}
\usepackage{graphicx}% Include figure files
\usepackage{dcolumn}% Align table columns on decimal point
\usepackage{bm}% bold math
\usepackage{xcolor}

\begin{document}

\preprint{APS/123-QED}

\title{Concept Study of a Storage Ring-based Gravitational Wave Observatory:\\Gravitational Wave Strain and Synchrotron Radiation Noise}

\author{Thorben Schmirander}%
\email{tschmira@physnet.uni-hamburg.de}
\affiliation{%
	Institute for Experimental Physics, Universit\"at  Hamburg, Luruper Chaussee 149, 
22761 Hamburg, Germany
}
\affiliation{Hamburg Observatory, Universit\"at  Hamburg, Gojenbergsweg 112, 
21029 Hamburg, Germany
}
\author{Velizar Miltchev}
\affiliation{%
	Institute for Experimental Physics, Universit\"at  Hamburg, Luruper Chaussee 149, 
22761 Hamburg, Germany
}

\author{Suvrat Rao}
\affiliation{Hamburg Observatory, Universit\"at  Hamburg, Gojenbergsweg 112, 
21029 Hamburg, Germany
}
\author{Marcus Br\"uggen}
\affiliation{Hamburg Observatory, Universit\"at  Hamburg, Gojenbergsweg 112, 
21029 Hamburg, Germany
}
\author{Florian Gr\"uner}
\affiliation{%
	Institute for Experimental Physics, Universit\"at  Hamburg, Luruper Chaussee 149, 
22761 Hamburg, Germany
}
\author{Wolfgang Hillert}
\affiliation{%
	Institute for Experimental Physics, Universit\"at  Hamburg, Luruper Chaussee 149, 
22761 Hamburg, Germany
}

\author{Jochen Liske}
\affiliation{Hamburg Observatory, Universit\"at  Hamburg, Gojenbergsweg 112, 
21029 Hamburg, Germany
}

\date{\today}

\begin{abstract}
This work for the first time addresses the feasibility of measuring millihertz gravitational waves (mHz GWs) with a storage ring-based detector. While this overall challenge consists of several partial problems, here we focus solely on quantifying design limitations imposed by the kinetic energy and radiated power of circulating ions at relativistic velocities. We propose an experiment based on the measurement of the time-of-flight signal of an ion chain. One of the dominant noise sources inherent to the measurement principle for such a GW detector is the shot noise of the emitted synchrotron radiation. We compute the noise amplitude of arrival time signals obtained by analytical estimates and simulations of ions with different masses and velocities circulating in a storage ring with the circumference of the Large Hadron Collider (LHC). Thereby, we show that our experiment design could reduce the noise amplitude due to the synchrotron radiation in the frequency range \({10^{-4} - 10^{-2} \;\text{Hz}}\) to one or two orders of magnitude below the expected GW signals from of astrophysical sources, such as super-massive binary black holes or extreme mass-ratio inspirals. Other key requirements for building a working storage ring-based GW detector include the generation and acceleration of heavy ion chains with the required energy resolution, their injection and continued storage, as well as the detection method to be used for the determination of the particle arrival time. However, these are not the focus of the work presented here, in which we instead concentrate on the definition of a working principle in terms of ion type, kinetic energy, and ring design, which will later serve as a starting point when addressing a more complete experimental setup.
\end{abstract}

\maketitle

\section{\label{sec:intro}Introduction}
GWs from the inspiral, ringdown and merging phases of super-massive binary black holes (SMBBH) or extreme mass-ratio inspirals (EMRI) arrive at Earth with millihertz frequencies \cite{Schmidt2020}. Measuring these signals will help us understand the masses and spins of super-massive black holes \cite{AmaroSeoane2007}, and provide insight into strong-field effects of general relativity \cite{Berti2006} and early galaxy formation \cite{Ju2000}. The laser interferometer space antenna (LISA) is currently developed to reach low noise levels in the frequency band of EMRI and SMBBH, and is planned to be operational by the next decade \cite{Vallisneri2009, Bailes2021}. It will be the first GW detector to bridge the frequency gap between pulsar timing arrays at very low frequencies \cite{Antoniadis2023} and frequencies of tens to hundreds of Hertz to which ground-based detectors such as LIGO are sensitive \cite{Weiss2018, Baker2019}.

Storage rings were already considered as a measurement device for GWs five decades ago \cite{Braginsky1977,ZerZion1998, VanHolten1999}, but the spatial deformations of the beam orbit due to the GW strain encoded in a beam position monitor signal were found to be proportional to the GW strain \(h^2\) \cite{Ivanov2021}, making it technically infeasible to measure such a signal. Recently, however, it has been noted that the velocity of a circularly traveling particle depends only linearly on the GW strain \(h\) and hence could be captured by a time-of-flight measurement \cite{Rao2020}, at least in principle. 

However, as any accelerated charge, circulating ions radiate off energy in the form of photons, which carry momentum and therefore lead to noise in the corresponding circulation time. For heavier ions, the impact of such a momentum kick on its velocity is smaller, compared to electrons or protons, making them preferable in such an experiment. In any case, measuring these small time differences with particles moving at very high energies will be experimentally challenging. For example, generating heavy ions, accelerating them to relativistic speeds and injecting them into a storage ring with extremely high temporal precision and low energy deviation will be a difficult task. Another difficulty will be finding an extremely precise, low-noise measurement procedure for actually determining the time-of-flight signal of the circulating particles in the storage ring. In this work we focus entirely on a numerical design study with the goal of minimizing the synchrotron radiation noise affecting the expected GW signal, which is already a complex problem in itself. Questions of the ion generation and acceleration as well as the time-measurements will be treated in future studies, for which the present work will provide valuable input. Hence this is the starting point of a complex design study.

Storage rings are ground-based detectors and gravitational gradient noise is much more prominent on Earth than in space, so that this poses less of a problem for space-based detectors, such as LISA \cite{Harms2013, Bailes2021, Martynov2016}. Since storage rings are usually designed for particle physics applications or as a source for synchrotron photons \cite{Shin2021}, they have to be repurposed to possibly measure GWs with low detection noise. This requires understanding and mitigating various noise sources, such as fluctuations in the magnetic field strength, temperature variations and vibrational noise. As a first step we focus purely on the influence of the GW on the time-of-flight signal of the circulating ions, as well as the quantum fluctuations of the emitted synchrotron radiation, and defer the other experimental challenges and noise sources to future work.
In this paper we attempt to answer whether in principle the time-of-flight signal of a particle circulating in a storage ring could be sensitive to strain from GWs and what requirements on the operation mode, particle type and setup need to be fulfilled in order to allow a mHz GW to surpass the noise from synchrotron radiation.

The paper is structured as follows: in Sec.\ \ref{sec:force}, we derive the Newtonian force in a coordinate system following the reference orbit, which is often referred to as Frenet-Serret coordinates, as a consequence of an incident GW. In Sec.\ \ref{sec:synchrad}, we discuss the impact of the shot noise caused by the synchrotron radiation on the traveling particle, as well as the numerical implementation of the 3D photon radiation pattern and the expected average deviation of the round trip time. In light of these findings, we propose an experimental setup in Sec.\ \ref{sec:prop_experiment}, that should allow to sufficiently reduce the photon shot noise imprinted on the time-of-flight signal to facilitate the measurement of a GW signal using the arrival time of the circulating particles. In Sec.\ \ref{sec:results}, we discuss how the GW strain could be reconstructed from the results of numerical computations. The power spectral density of the shot noise is computed and compared to the characteristic strain of mHz GW sources, from which an optimal frequency window for the operation of a storage ring-based GW detector is obtained. Finally, we summarise our findings in Sec.\ \ref{sec:summary}.

\section{\label{sec:force}Newtonian Force in Storage Ring Frame of Reference}
The starting point of our investigation is the Newtonian force of a GW which acts on the circulating particles, described in a coordinate system following the design orbit of the particle which is commonly used in accelerator physics. In the following, we derive the equations of motion for a particle traveling through different elements of a storage ring (drift space, sector magnet, quadrupole magnet), while being subject to an external, slowly varying force due to an incoming GW. Some assumptions must be made in order to reduce the complexity of the problem. Since mHz GWs vary slowly on a time scale of hours, the effective GW strain \(h_{\theta\phi\psi}(t)\) in the frame following the orbit of the design particle is assumed to be discretized into time intervals of length \(\Delta t\) and thus the GW force at a given storage ring section, parametrized by the angle \(\alpha\), is assumed to be constant in time for a number of revolutions \(n\) of the particle, covering \(\Delta t \approx 0.5\)s. The storage ring dimensions are always assumed to be much smaller than the wavelength of the GW, and thus the phase of the GW is assumed to be constant over the whole ring for at least one whole revolution of the particle. 

\subsection{Longitudinal Force}

\begin{figure}[t]
  \includegraphics[width = 0.5\textwidth]{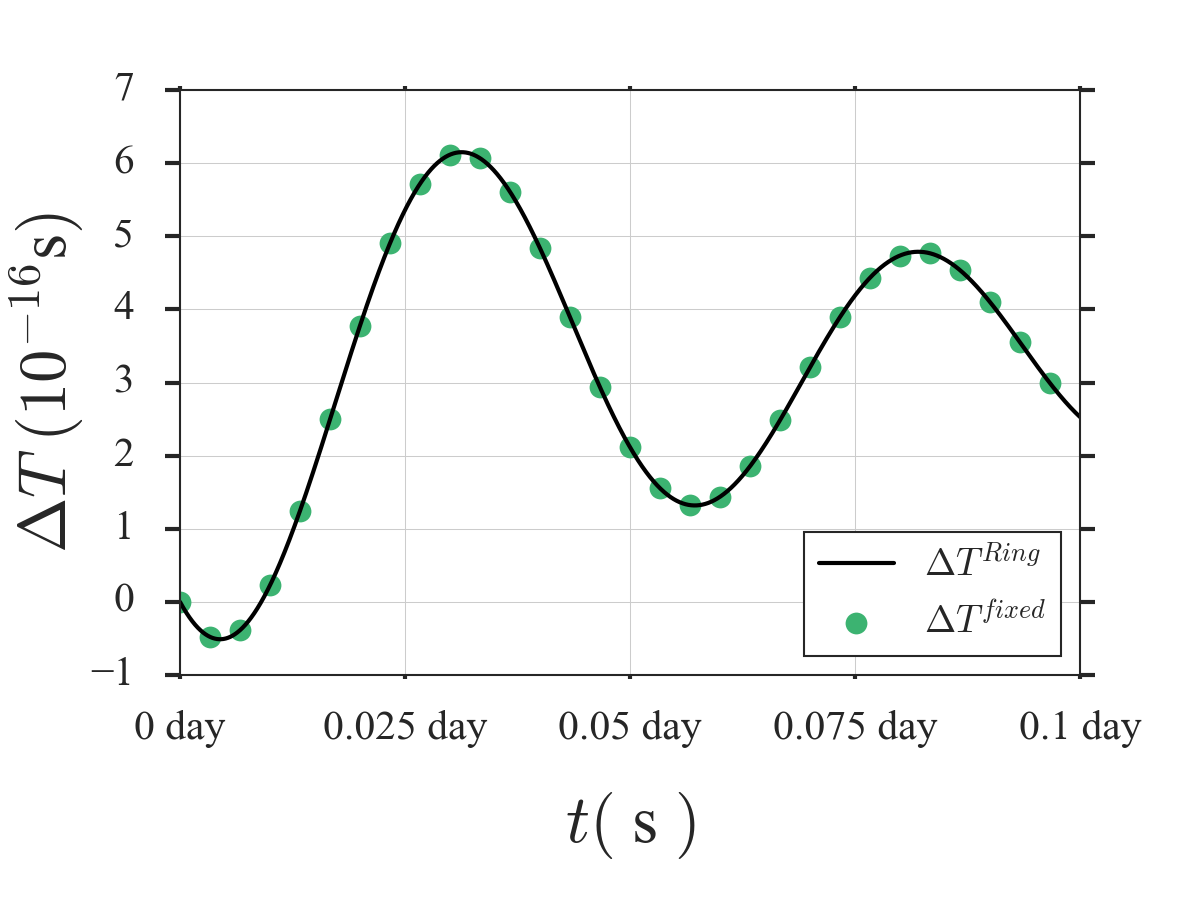}
  \caption{\label{fig:signal_fit} An example GW signal as obtained from the timing deviation of a storage ring on Earth with a radius of \(120\) m and for a proton with \(\beta=0.9\) over a time interval of 2.5 hours \cite{Rao2020}. The inspiral of two non-spinning objects, each with \(10^6\,M_\odot\), an initial separation of \(0.32\) AU, at redshift \(z=0.2\) and at a distance from Earth of \(990\) Mpc is modeled using Post-Newtonian analysis. The analytical prediction (green dots) and numerical computation for a single particle subject to an external force (black line) are directly related via the slip factor, which encodes the storage ring geometry, and a relativistic correction, see Eq.\ \ref{relation_signal_strength_main}.}
\end{figure}

For relativistic motion, an acceleration \(c \dot{\beta_\parallel}\) parallel to the direction of motion of a particle with rest mass \(m\) and velocity \(v_{0} = c \beta_\parallel\) leads to a force \(F_\parallel\) given by
\begin{align}
F_\parallel & = mc\dot{\beta_\parallel}\gamma^3,
\label{longitudinal_force_0}
\end{align}
where \(\gamma = 1/\sqrt{1-\beta_\parallel^2}\), and \(c\) is the speed of light  \cite{Rafelski2019}. Since the storage rings discussed in this work have large circumferences, they are assumed to be perfect circles for computing the force of the GW strain in the ion's rest frame. In this case, storage rings have cylindrical symmetry and the spacetime metric is given by \cite{Rao2020}, 
\begin{align}
ds^2 &=-c^2dt^2 + \left(1 + h_{\theta\phi\psi}(t,\alpha)\right)R^2d\alpha^2,
\end{align}
where \(h_{\theta\phi\psi}(t,\alpha)\) is the GW strain as measured around a ring with radius \(R\) at an angle \(\alpha\). In the following, we will adopt the notation of \cite{Rao2020} where required. The gravitational strain of the incident GW is given by
\begin{align}
h_{\theta\phi\psi}(t,\alpha) &= h_+(t)\left(f^+_s \sin^2{\alpha} + f^+_c \cos^2{\alpha} + f^+_{sc} \sin{2\alpha}\right)\nonumber \\
&+ h_{\times}(t)\left(f^{\times}_s \sin^2{\alpha} + f^\times_c \cos^2{\alpha} + f^\times_{sc} \sin{2\alpha}\right),
\end{align}
with the two GW polarizations denoted by indices (\(\times\)) and (\(+\)), respectively, and where the coefficients \(f^{+/\times}_{s/c/sc}\) reflect the Earth's rotation and are given by expressions of the Euler angles \(\theta, \phi\) and \(\psi\) \cite{Rao2020}. This leads to three geodesic equations, from which the longitudinal acceleration can be derived, as
\begin{align}
\frac{d^2 l}{d\tau^2} + \frac{1}{1+h_{\theta\phi\psi}(t)}\frac{dh_{\theta\phi\psi}(t)}{dt}\left(\frac{dl}{d\tau}\right)\left(\frac{dt}{d\tau}\right) &= 0,
\end{align}
which can then be transformed into
\begin{align}
\frac{d^2l}{dt^2} &= - \frac{1}{1 + h_{\theta\phi\psi}(t)}\frac{dh_{\theta\phi\psi}(t)}{dt} v_0 \label{geodesic_equation}.
\end{align}
Here, \(l\) denotes the path arc length around the ring and \(\tau\) the eigentime of the particle \cite{Rao2020}. Assuming that the GW does not change the velocity of the particle too much, the parallel velocity \(v_\parallel\) can be replaced by the initial velocity \(v_0\), such that with \(c \dot{\beta_\parallel} = a_\parallel(t) = \frac{d^2l}{dt^2}\), the longitudinal acceleration leads to the force of a GW on a circularly traveling particle of
\begin{align}
F_\parallel(t) = -m\gamma^3\frac{\dot{h}_{\theta\phi\psi}(t)}{1 + h_{\theta\phi\psi}(t)}v_0,
\label{longitudinal_force}
\end{align}
by inserting Eq.\ \ref{longitudinal_force_0} into Eq.\ \ref{geodesic_equation}. Here the dependence on \(\alpha(t)\) has been absorbed into the time-dependence via \(\alpha(t):= v_0 t/R\).

\subsection{Relating Storage Ring Signal to Signal in Ring with Fixed Radius}
A key assumption of the idealized model discussed by Rao et al.\ \cite{Rao2020} is that the radius of the trajectory of the traveling particle is constant while it is subject to the GW strain. In a real storage ring, however, this assumption is invalid, as the de- or increased momentum of a particle due to an external force also leads to a change of the relativistic mass \(m \gamma\), causing a change in the radius of curved paths in magnetic fields and thus a non-linear change of the circulation time. This is encoded in the slip factor \(\eta\), and the transformation is derived in detail in the Appendix \ref{app:relating_signals}. Here, we just summarise the main findings. The Newtonian acceleration from a GW in the storage ring frame of reference is approximated as
\begin{align}
a_\parallel(t) & \approx -\dot{h}_{\theta\phi\psi}(t)v_0, 
\end{align}
because \(h_{\theta\phi\psi}(t) \ll 1\). In a ring with fixed radius, the time integral of the longitudinal acceleration \(a_\parallel\) directly leads to \(\Delta v(t)\), which integrated over time again leads to the longitudinal separation
\begin{align}
\Delta l^{\text{fixed}}(t) &= -v_0\int_0^t\left( h_{\theta\phi\psi}(t') - h_{\theta\phi\psi}(0)\right)dt'.\label{runtime_deviation}
\end{align}
Up to a factor, this result is similar to the fully relativistic expression obtained by Rao et al.\ \cite{Rao2020}. By correctly accounting for the influence of the particle momentum on the circulation, one can relate the real storage ring signal \(\Delta T(t)^\text{Ring}\) to the signal expected in a ring with fixed radius, \(\Delta T(t)^\text{fixed}\), via
\begin{align}
\Delta T(t)^\text{fixed}=  &- \frac{1}{\eta}\frac{1}{\gamma^2}\left(1 - \frac{v_0^2}{2c^2}\right) \Delta T(t)^\text{Ring} \nonumber \\
&- \left(1 - \frac{v_0^2}{2c^2}\right) \overline{h_{\theta\phi\psi}(t_0,\alpha_0)} t, \label{relation_signal_strength_main}
\end{align}
where the time dependence has been included explicitly and \(h_{\theta\phi\psi}(0)\) has been replaced with the average \(\overline{h_{\theta\phi\psi}(t_0,\alpha_0)} = \frac{1}{N}\sum_{i\in\text{Ring}}^Nh_{\theta\phi\psi}(t_0,\alpha_i)\) over the positions \(\alpha_i\) for all \(N\) ring elements (magnets and drift spaces). The last term in Eq.\ \ref{relation_signal_strength_main} originates from the initial spacetime strain, by which the storage ring metric differs from flat space.

An example of \(\Delta T(t)^{\text{fixed}}\) is shown in Fig.\ \ref{fig:signal_fit}, where an example signal of a mHz GW is used to compute the predicted time deviation using the numerical procedure described in \cite{Rao2020}. This is compared to the result of an ion tracked in a storage ring and the resulting noise-free time deviation transformed by Eq.\ \ref{relation_signal_strength_main}, coinciding with the result for a fixed ring. The result of a numerical study using particle tracking in two example storage rings is shown in Appendix \ref{app:relating_signals_numerics}.  

\subsection{Numerical Methods}

\begin{figure}[t]
  \includegraphics[width = 0.45\textwidth]{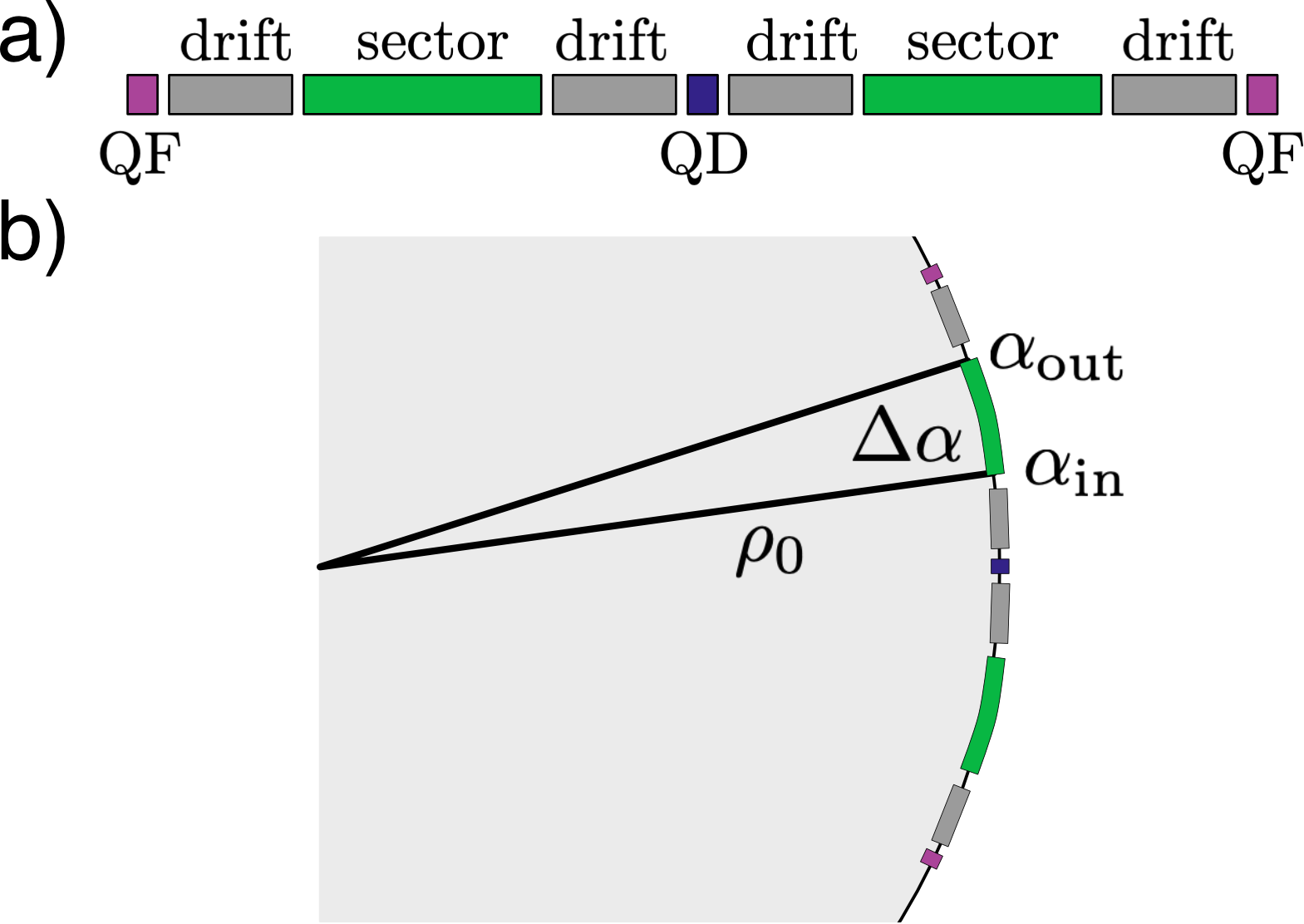}
  \caption{\label{fig:unit_cell_schematics}a) Schematic of the FODO magnetic unit cell used in this work, consisting of sector dipole and quadrupole magnets. Due to symmetry, each unit cell starts and ends with a radially focusing quadrupole magnet (QF) and has a radially defocusing quadrupole magnet (QD) in the center, separated by drift spaces and sector magnets. b) For the computation of the GW strain the center of each ring element is determined using polar coordinates with angle \(\alpha\) and bending radius of the sector magnets \(\rho_0\).}
\end{figure}

\begin{figure}[b]
  \includegraphics[width = 0.5\textwidth]{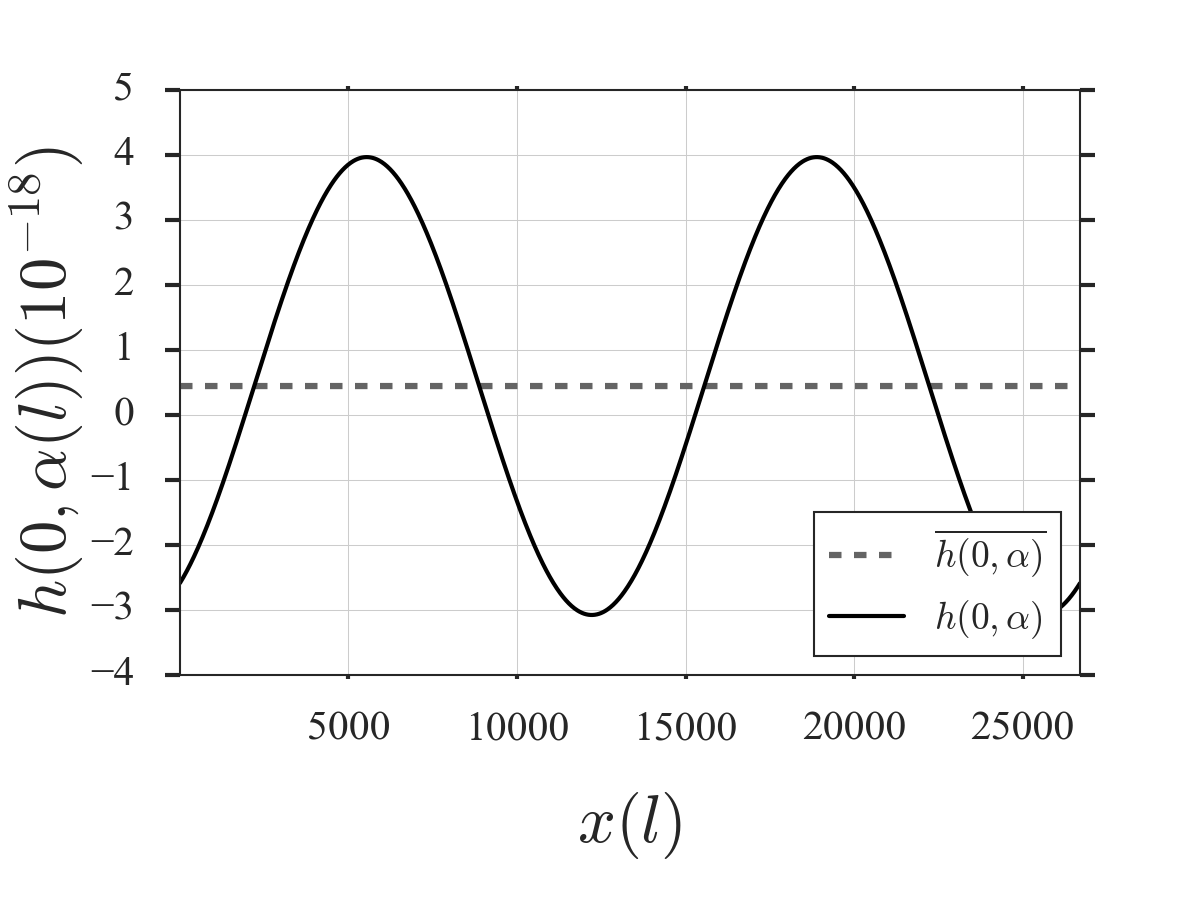}
  \caption{\label{fig:strain_around_ring} The GW strain of the incoming GW at the beginning of the measurement in the storage ring frame of reference using cylindrical coordinates. The angle \(\alpha(l)\) is parametrised by the path length \(l\). The average GW strain for a single turn is shown as dashed line.}
\end{figure}

A particle in a storage ring is tracked via transfer matrices, which correspond to the different elements of the magnetic lattice and which define the storage ring geometry \cite{Hinterberger2008}. From these transfer matrices, the momentum compaction factor \(\alpha_p\) and the slip factor \(\eta\) can be computed \cite{Hinterberger2008}, see also Appendix \ref{app:relating_signals}, which mainly encode the optical characteristics of the storage ring important for the timing deviation in this work. To include an external force, originating from a GW passing through the storage ring on Earth, the standard equations of motion for the six-dimensional phase space of the magnetic lattice have to be modified, which is presented in Appendix \ref{app:matrices}. Since the impact of the external force changes along the ring, a particle has to be tracked through each ring element individually and cannot be computed by using the transfer matrix for a whole unit cell or the entire ring.

For simplicity, all storage rings discussed in this work will be assumed to be comprised of regular FODO cells, consisting of regular sector dipole magnets, drift spaces and quadrupole magnets \cite{Hinterberger2008}, with geometry shown in Fig.\ \ref{fig:unit_cell_schematics} a). As the dimension of a single ring element is small compared to the ring circumference, the GW force within each ring element (located at a section starting at \(\alpha_\text{in}\) and ending at \(\alpha_\text{out}\)), see Fig.\ \ref{fig:unit_cell_schematics} b), is assumed to be constant in polar angle \(\alpha\) of the ring, as well as time \(t\), where \(\alpha\) is taken in the middle of said element, such that \(\alpha = \alpha_\text{in} + \frac{\alpha_\text{out}-\alpha_\text{in}}{2}\). An example of an effective force along the ring is shown in Fig.\ \ref{fig:strain_around_ring} for an example signal of an SMBBH inspiral. The average strain, and thus average force, along the ring is non-zero, indicated by the dashed curve in the plot.

The longitudinal deviation of a particle due to a GW is very small and will therefore be considered here only up to first order. Thus, despite a finite momentum deviation of a particle \(\Delta p_0\) from the reference particle with \(p_0\) and \(v_0\), this momentum, used in the computations of the force and shifts, will be approximated as \(p = p_0 + \Delta p_0 \approx p_0\). Depending on the ring size, a particle takes \(10^3 - 10^6\) turns within \(0.5\) s, during which the GW strain within each individual element of the ring is assumed to be a constant in time, defining a time step in the computation. 

The computed results for the timing deviation of a particle circulating in a storage ring are then transformed using Eq.\ \ref{relation_signal_strength_main}. The results obtained with different storage ring geometries and particle energies are thus rendered comparable, which will be important when comparing noise amplitudes.

\section{\label{sec:synchrad}Synchrotron Radiation with Single-Photon Resolution}
In this section we compute the noise in the arrival times of the circulating ions that arises from the stochastic nature of the synchrotron emission radiated by these particles. Synchrotron radiation is usually characterized by the average emitted power, but the particle dynamics of ions circulating in a ring depend on the individual momenta of the emitted photons, as well as the rate of the photon emission, both of which fluctuate around the classical emitted power \cite{Saldin1996}. In addition, the synchrotron photons are emitted with some spatial distribution. All of these effects lead to a velocity-dependent arrival time noise.

In the following, we first recall the photon statistics required for particle tracking \cite{Jackson1962, Sands1980}, and then discuss a numerical implementation of how the photon momenta, such as the average power and the synchrotron spectrum, are computed from the classical results. We assume that synchrotron radiation is only emitted in the curved sections of the ring, i.e.\ within the sector magnets. The equations of motion for a particle emitting a number of photons in a sector magnet are derived in Appendix \ref{app:synchrad}.
\subsection{Photon Statistics}
The power for a circular acceleration of a particle on a trajectory with radius \(\rho\) in appropriate units is given by \cite{Jackson1962}
\begin{align}
P_\gamma &= \frac{2}{3}r_c m c^3\frac{\beta^4 \gamma^4}{\rho^2},
\label{emitted_power}
\end{align}
where \(r_c = \frac{e^2}{4\pi \varepsilon_0 m c^2}\). The critical energy is given by
\begin{align}
u_c &=\frac{3}{2}\frac{\hbar c \gamma^3}{\rho}
\end{align}
and the average number of photons emitted per unit time is
\begin{align}
\dot{N} &= \frac{15 \sqrt{3}}{8} \frac{P_\gamma}{u_c}
\label{full_photon_number}
\end{align}
\cite{Sands1980}.
As mentioned above, we assume that photons are only emitted in the sector magnets. In this case, the expected number of emitted photons per turn is determined by the time spent in the sector magnets during one revolution, such that
\begin{align}
  \dot{N}_\text{exp} &= \dot{N} \frac{2\pi \rho}{\beta c}.
\label{expected_photon_number}
\end{align}
The photon number in Eq.\ \ref{expected_photon_number} is Poisson-distributed \cite{Jaeschke2020}.
To obtain the total energy, the universal synchrotron spectrum, 
\begin{align}
S(\xi)&=\frac{9\sqrt{3}}{8\pi}\xi \int_\xi ^\infty K_{5/3}(x)dx
\label{universal_synchrotron_function}
\end{align}
is used, where \(K_{5/3}(x)\) is the modified Bessel function of the second kind and where \({\xi = u/u_c}\) is the rescaled energy \(u\). The spectral photon density at this energy is given by
\begin{align}
n(\xi)&=\frac{P_\gamma}{u_c^2}\frac{1}{\xi}S(\xi)
\label{photon_number}
\end{align}
\cite{Sands1980}. By using Eq.\ \ref{photon_number}, a probability distribution is defined due to the fact that
\begin{align}
\int_0^\infty u_c \frac{n(\xi)}{\dot{N}_\text{exp}}d\xi = 1.
\label{probability_photons}
\end{align}
For numerical purposes, the values of Eq.\ \ref{probability_photons} are pre-computed on the interval \(\xi \in [10^{-14},20]\)\footnote{This interval is chosen to approximate \([0,\infty)\).} and a cumulative probability distribution is defined and then inverted, to draw photon energies \(u = u_c \xi\) with the appropriate probabilities. The expected photons emitted per revolution \(\dot{N}_\text{exp}\), together with the expected energy per photon \(\langle u \rangle = \frac{8}{15 \sqrt{3}}u_c\) lead to a mean effective emitted power of
\begin{align}
\langle P_\text{eff} \rangle = P_\gamma \frac{2\pi \rho}{L_\text{Ring}} = \dot{N}_\text{exp}\frac{\beta c}{L_\text{Ring}} \langle u \rangle,
\end{align}
where \(L_\text{Ring}\) is the circumference of the entire storage ring.

\subsection{\label{sec:radiation_pattern}3D radiation pattern}
In order to compute the probability of the photon emission angles, a function proportional to the 3-dimensional pattern for photons emitted by accelerated charges \cite{Jackson1962} is used, where the polar angle is given by \(\theta\) and the azimuthal angle by \(\phi\). It is denoted
\begin{align}
\vec{F}(\theta,\phi) =& \nonumber \\
&\frac{\left(1-\beta \cos{(\theta)}\right)^2 - \left(1-\beta^2\right)\sin{(\theta)}^2\cos{(\phi)}^2}{\left(1-\beta \cos{(\theta)}\right)^5} \nonumber\\
&\times
\begin{pmatrix}
\sin{(\theta)}\cos{(\phi)}\\
\sin{(\theta)}\sin{(\phi)}\\
\cos{(\theta)}
\end{pmatrix},
\end{align}
and as it is used to weight the probability for a photon to be emitted at a given angle, the distribution is normalized by \(|\vec{F}(0,0)|\). In order to obtain a collection of angles \(\{\theta, \phi\}\), which is weighted by this emission pattern, first, a number of random angles \(\{\Bar{\Bar{\theta}}, \Bar{\phi}\}\) is drawn from the interval \([0, \pi]\times [0, 2\pi]\). These angles do not represent an equal distribution of points on a spherical surface, but rather on flat space. Thus, all angles \(\Bar{\Bar{\theta}}\) have to be re-weighted by a procedure described in Appendix \ref{appsup:inverse_function}. Each pair of angles from the resulting set \(\{\Bar{\theta}, \Bar{\phi}\}\) is then assigned a third random number \(f_i \in [0,|\vec{F}(0,0)|]\). Using simple rejection sampling, each triplet of numbers \(\{\Bar{\theta}_i, \Bar{\phi}_i, f_i\}\) determines  the assignment of the angles to the collection of the weighted \(\{\theta, \phi\}\) via the condition \(f_i \leq |\vec{F}(\Bar{\theta}_i, \Bar{\phi}_i)|\). For every photon the emission will then happen in direction of a unit vector with a pair of angles from this collection, such that the wave vector of the \(i\)th photon emitted with energy \(u_i\) is given by \(\vec{k}_i = \frac{u_i}{c \hbar} \left(\sin{(\theta_i)}\cos{(\phi_i)}, \sin{(\theta_i)}\sin{(\phi_i)}, \cos{(\theta_i)}\right)\).

\subsection{\label{sec:synchrotron_time_of_flight}Impact on Time-of-Flight Signal}
For the numerical implementation, the number of photons per revolution is computed via a Poisson distribution with a mean given by Eq.\ \ref{expected_photon_number}. As all FODO cells are comprised of identical magnetic elements, each photon is assigned to one sector magnet in one specific magnetic unit cell, with equal probability. The equations of motion for the particle are derived in Appendix \ref{appsup:eom}. By using the integrand of Eq.\ \ref{probability_photons} as probability distribution to sample the synchrotron spectrum, the energy of each photon is obtained and then one pair of angles from the collection \(\{\theta, \phi\}\) defined in the previous section is assigned. The recoil imprinted on the moving ion by each of these photons is then used in the equations of motion to track the particle in the storage ring.\\
The expected timing deviation of the particle due to the loss of energy is computed as the average effect from the radiation, because every emitted photon transfers the momentum \({\Delta p_\text{ion} = -\hbar k_\text{photon}}\), but the force from a number of emissions after the time \(\Delta t\) is
\begin{align}
\frac{dp}{dt} = -\frac{\sum_i \hbar k_i}{t} = m \gamma^3 a,
\end{align}
where \(\sum_i \hbar k_i = \frac{\langle P_\text{eff} \rangle t}{c}\) and \(m\) is the particle mass, leading to the acceleration
\begin{align}
a &=\frac{\langle P_\text{eff} \rangle}{c m \gamma^3}.
\end{align}
It follows for the time deviation due to the average slow down from synchrotron emission
\begin{align}
\Delta T_0 &= t^2 \frac{\langle P_\text{eff}\rangle}{2 p_0 c \gamma^2},
\label{time_deviation_synch_rad}
\end{align}
when the total power is emitted in the longitudinal direction. Due to the 3D radiation pattern, not all photons are emitted in an exactly longitudinal direction, but are instead emitted in all directions with some angular probability distribution. The net slow-down is given by the average photon emission in the longitudinal direction, because the emission pattern has cylindrical symmetry with respect to this axis, such that the other two components cancel out on average. Furthermore, by employing numerical studies, the radial direction of emission are shown to have an effect on the time-of-flight signal several orders of magnitude smaller than the longitudinal direction. These considerations lead to the weighting factor
\begin{align}
S_0 &= \frac{1}{N}\sum_i^N \cos{(\theta_i)},
\label{net_factor_longitudinal}
\end{align}
with \(N\) being the total number of emitted photons, such that the total timing deviation due to the synchrotron radiation is given by 
\begin{align}
  \Delta T &= \Delta T_0 S_0.
\end{align}
A typical value for \(\beta = 0.32\) is \(S_0 \approx 0.32\), which will be suitable for lowering the shot noise of the synchrotron radiation, as discussed in the following sections. As can be seen from Eq.\ \ref{time_deviation_synch_rad}, the average timing deviation increases with the duration of the measurement as \(t^2\). This time-dependent average value can be used to predict the expected arrival time of the circulating ions, leaving only the uncertainty of their revolution time due to the stochastic nature of the synchrotron photon emission. 

\section{\label{sec:prop_experiment}Proposed Experimental Setup}
In this section we discuss the particle type and operation mode of a storage ring that are required to potentially detect the signature of a GW strain encoded in the arrival time signal of the circulating particles.

\subsection{Storage Ring Operation}
\begin{figure}[t]
  \includegraphics[width = 0.4\textwidth]{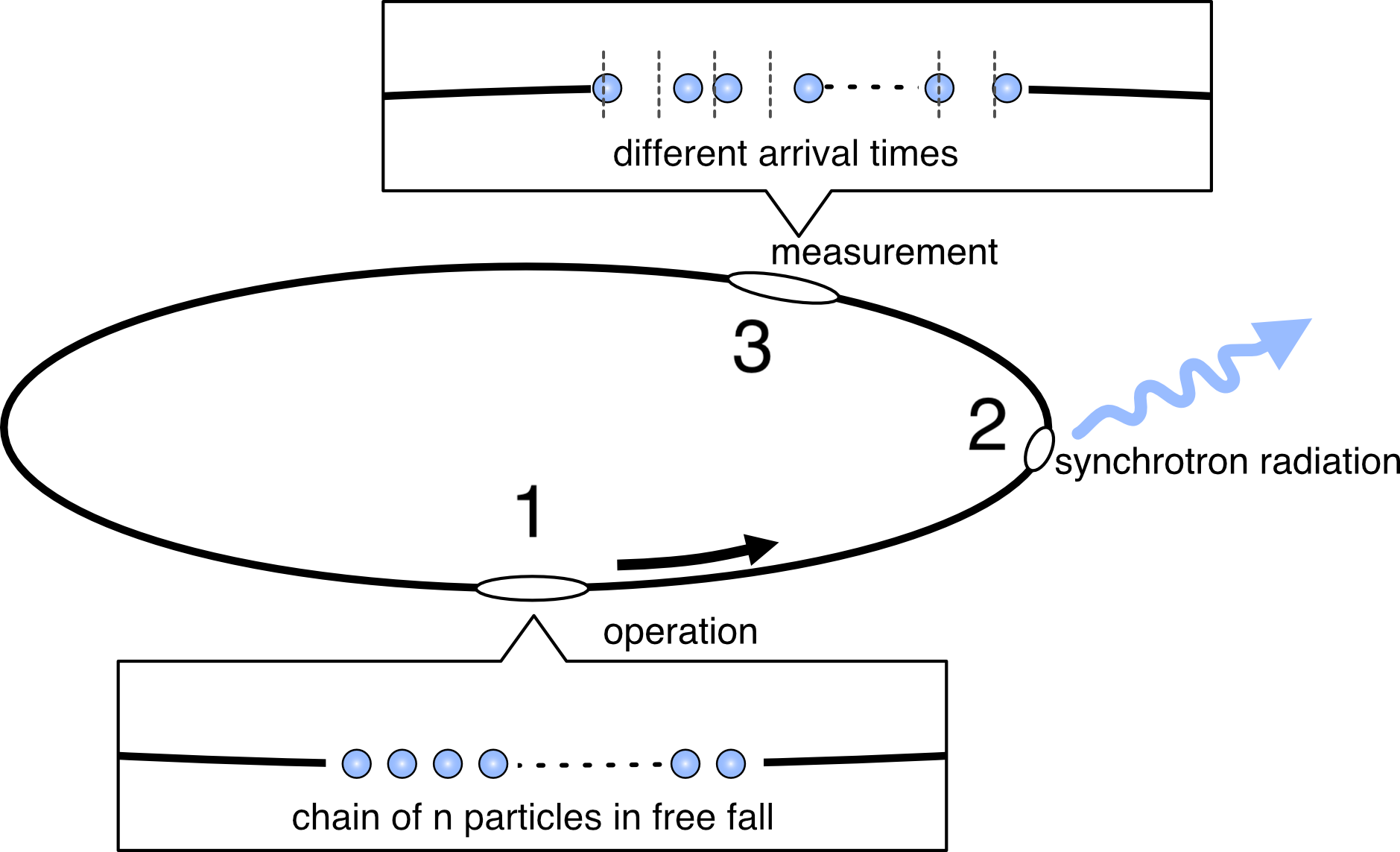}
  \caption{\label{fig:exp_setup} An ion chain is injected into the storage ring and all positions at the initialisation time are known (1). While circulating, synchrotron radiation is emitted within the sector magnets (2). The quantum fluctuations of the emission cause each ion to exhibit a different offset from the expected arrival time (dashed lines) after circulating in the ring for many revolutions (3).}
\end{figure}

The proposed experimental setup consists of a chain of \(n\) particles, which are injected into a storage ring at a certain energy, without the use of a radio frequency cavity, as opposed to common practice in storage rings \cite{Saha2016}. In such a setup the free fall of the particles in the azimuthal direction is maintained, which is depicted in Fig.\ \ref{fig:exp_setup}. This chain of ions is separated longitudinally well enough, such that Coulomb interactions among the ions can be neglected. Due to the slow variation of the GW strain over time, however, the force acting on these particles can be regarded as identical. In the sector magnets, synchrotron radiation is emitted, which has the same mean value for all particles, but individual fluctuations cause a jitter in the arrival time. If we average the arrival times of the particles, and if the number of particles is large enough, these fluctuations should mutually cancel. Then, the quadratic slow-down due to the average energy emission can be subtracted and only the uncertainty of the mean value, as obtained from the \(n\) ion arrival times, remains -- with enough accuracy to possibly infer the strain of a passing GW. 

For measuring the arrival times, an idealised measurement device with sub-fs precision is assumed in this feasibility study, whereas a detailed assessment of such a detection setup will be discussed in future publications. Here, the focus will be put on reducing the synchrotron emission noise and finding a first working principle. In case of a laser measurement technique, the measurement procedure will probably have to be fine-tuned to the particle type, and its internal states, as well as its velocity. Furthermore, the circulation time of the particle must be synchronized with the ticking of a precise atomic clock \cite{Rao2020}. For the characterisation of the synchrotron radiation noise, however, the exact mechanism of arrival time measurement is irrelevant\footnote{The measurement uncertainty due to such a device will be the topic of a future study.}, because the synchrotron noise originates entirely from the photon shot noise and therefore cannot be mitigated by any technical device. If the particles circulate in the ring for longer, the arrival time uncertainty will build up and will eventually surpass the time delay caused by the GW. In order to avoid this, after a suitable time period \(\Delta t\), the \(n\) ions have to be dumped and \(n\) new particles must be injected into the ring. Immediately after injection, these new particles exhibit no time delay originating from the synchrotron emission or the GW signal, such that their initial circulation time practically corresponds to the unperturbed revolution time, and a time delay will only slowly start to build up. The uncertainty of the arrival time of these new ions is therefore also very small immediately after injection, such that a suitable fitting algorithm may determine the GW form with much more accuracy from the data taken shortly after injection compared to the data taken when the ions have been circulating for longer periods of time.

\subsection{Particle Velocity}
The particle velocity \(\beta c\) is clearly an important operational choice, which will affect the experiment on many different levels. For example, depending on the instrumentational choice for the arrival time measurement, the measurement noise may well depend on \(\beta\). Eventually we will want to determine the best value of \(\beta\) taking all aspects of the experiment into consideration, including all of the known noises sources. However this global optimization is beyond the scope of this paper.

Here, we confine ourselves to an optimization of \(\beta\) with respect to the synchrotron noise while taking into consideration some general operating constraints of the storage ring, such as realistic magnet strengths and residual gas collisions.

For large ion velocities the uncertainty in the arrival times of the particles is dominated by both the Poisson uncertainty of the synchrotron emission rate (cf.\ Eq.\ \ref{expected_photon_number}) and the photon energy spread (cf.\ the integrand of Eq.\ \ref{probability_photons}). In contrast, for smaller velocities, the synchrotron emission is less beamed and more spatially homogeneous, such that the arrival time noise is instead dominated by the uncertainty in the direction of the emission.

The ideal choice of ion velocity, where the lowest noise is expected, may thus be determined by simultaneously minimizing the photon energy fluctuation and the emission rate fluctuation, while also considering that higher velocities lead to relatively more photon emission in the forward direction whereas slower ions emit more homogeneously in all spatial directions. These considerations lead to an optimal value of \(\beta \approx 0.13\), below which the noise cannot be reduced substantially any further.

However, additional considerations prevent us from using this optimal choice of \(\beta\). 
For the storage rings discussed below (see Table \ref{tab:table1}), a \(\beta\) as low as \(0.13\) would require the sector magnets to be operated with a magnetic field below \(100 \text{ mT}\), which in turn would lead to additional noise sources, such as the Earth's magnetic field. In addition, our numerical studies have shown that the synchrotron radiation noise for \(\beta = 0.13\) is effectively not much lower than for \(\beta=0.32\), which in our opinion does not justify the potential increase of other noise sources.

Hence, we consider \(\beta = 0.32\) as the lowest realistic value and we will adopt this value in the analysis below. We will also consider \(\beta = 0.95\) in order to illustrate the impact of a larger \(\beta\) on the synchrotron radiation noise. These values may be regarded as the extreme ends of the feasible beta range to show that signals of SMBBH and EMRI sources could be detected in principle using our detection concept.

\subsection{Particle Type and Ring Dimensions}

The arrival time jitter for particles with masses on the order of protons has been found to exhibit too much noise, similar to particles with multiple charges, even for storage rings with large bending radii. As was discussed before, the impact of the carried-off photon momentum on a particle's velocity is smaller for a heavier ion, while the synchrotron emission power (Eq.\ \ref{emitted_power}) increases with the velocity and charge of a particle, but does not depend on its mass \cite{Jackson1962}. Therefore, from now on we only consider singly-charged heavy ions, which offer a small charge-to-mass-ratio and thereby lead to the smallest possible emitted power and minimal impact of the photon recoil on its velocity, simultaneously.

A candidate ion species should exist in large abundance and should be stable under acceleration and continued circulation in a storage ring for several hours. Therefore, complex molecules, unstable radioactive isotopes with short half lifes or ions, which can only be obtained in small quantities, are unsuitable candidates. For optimal results, the computations here are performed using either a singly-charged \(^{238}\text{U}\) ion or a singly-charged \(\text{U}_2\) molecule, consisting of two \(^{238}\text{U}\) atoms. Ions with similar weight, however, such as Te\(_2^{-}\) \cite{Snodgrass1989}, consisting of two \(^{130}\text{Te}\) atoms, could lead to very similar results. For \(\text{U}^{+/-}\) or \(\text{U}_2^{+/-}\) ions, the difference of one electron mass is negligible compared to the mass of the neutral ion, such that the mass of the neutral particle is used in the following for both cases. For particles much heavier than \(\text{U}_2\) it becomes technically infeasible to accelerate and store them using the field strengths of storage ring magnets used today. The chemical properties, especially the bond order, of the diuranium molecule \(\text{U}_2\) and the ionized \(\text{U}_2^-\), is an active topic of research \cite{Ciborowski2021, Gagliardi2005, Knecht2019}. Using laser vaporization techniques, however, the anions \(\text{U}^-\) \cite{Tang2021} and \(\text{U}_2^-\) \cite{Ciborowski2021} can be synthesized for the use in experiments, whereas other types of singly charged cationic uranium molecules with intermediate masses are experimentally accessible \cite{Marks2021} and could potentially be used, as well. For the computations discussed in the following, several example storage rings are designed, stable for \(\text{U}^{+/-}\) and \(\text{U}_2^{+/-}\) ions at different velocities. Each of these rings has the circumference of the Large Hadron Collider (LHC) of \(26.7\) km and consists of FODO cells and sector magnets only, with the dimensions as stated in Tab.\ \ref{tab:table1} and featuring realistic magnetic field strengths. In an imperfect vacuum, recombination of the positive ions \(\text{U}^{+}\) and \(\text{U}_2^{+}\) with free electrons could occur, rendering the particles useless for the experiment. However, currently, given a good enough vacuum, experiments using a countable number of electrons with lifetimes exceeding several minutes can be performed to study single particle dynamics \cite{Lobach2022, Romanov2021}. In addition, heavy ions with higher kinetic energies are less likely to recombine with the background gas as slower ions \cite{Xue2009}. For molecules, such as \(\text{U}_2^{+/-}\), also molecular vibrations could be excited upon passage of a sector magnet, and the magnetic fields might dissociate such a molecule into two single uranium ions. For measuring small distances, the spatial extensions of the heavy ions used in the experiment may become relevant. The bond length of the diuranium molecule is \(2.43 \text{ \normalfont\AA}\) \cite{ciborowskiSupplementaryMaterials2021} and a single uranium atom has a classical radius estimated to be \(174 \text{ pm}\) \cite{Slater1964}. For a particle velocity of \(\beta = 0.32\), however, assuming a required measurement resolution of \(0.3 \times 10^{-16}\text{ s}\), as estimated from Fig.\ \ref{fig:signal_fit}, a spatial measurement precision of around \(2.9 \text{ nm}\) for determination of the time-of-flight signal is required. Thus, we conclude that the particles \(\text{U}^{+/-}\) and \(\text{U}_2^{+/-}\) could in principle be used in our experiment design and would allow us to reach the required precision in the measurement of their arrival time.
\begin{table*}
  \caption{Parameters for the rings with circumference of \(26.7\) km, similar to the LHC, for \(\text{U}^{+/-}\) and for \(\text{U}_2^{+/-}\), operated at either \(\beta = 0.32\) or \(\beta = 0.95\). Each magnetic unit cell consist of radially focusing (QF) and defocusing (QD) quadrupole magnets with magnetic parameters \(k^{\text{QD}/\text{QF}}\), sector magnets with edge focusing and drift spaces, which have lengths \(l^\text{sector}\) etc.\label{tab:table1}}
  \begin{ruledtabular}
  \begin{tabular}{ccccccccccccc}
   &\(\beta\)&\(k^\text{QF}\)&\(k^\text{QD}\)&\(l^\text{QD}\)&\(l^\text{QF}\)&\(l^\text{sector}\)&\(l^\text{drift}\)&\(n_\text{cells}\)\\
   \hline
   \(\text{U}^{+/-}\)&0.32&\(0.891\frac{\text{T}}{\text{m}}  \cdot \frac{q}{p_0}\)&\(1.656\frac{\text{T}}{\text{m}}  \cdot \frac{q}{p_0}\)&\(3.250\text{m}\)&\(3.600\text{m}\)&\(47.10\text{m}\)&\(7.300\text{m}\)&\(200\)\\
   \(\text{U}_2^{+/-}\)&0.32&\(1.871 \frac{\text{T}}{\text{m}} \cdot \frac{q}{p_0}\)&\(3.478 \frac{\text{T}}{\text{m}} \cdot \frac{q}{p_0}\)&\(2.950\text{m}\)&\(3.400\text{m}\)&\(47.10\text{m}\)&\(7.500\text{m}\)&\(200\)\\
   \(\text{U}^{+/-}\)&0.95&\(9.801\frac{\text{T}}{\text{m}}  \cdot \frac{q}{p_0}\)&\(13.69\frac{\text{T}}{\text{m}}  \cdot \frac{q}{p_0}\)&\(3.000\text{m}\)&\(4.100\text{m}\)&\(47.10\text{m}\)&\(7.300\text{m}\)&\(200\)\\
   \(\text{U}_2^{+/-}\)&0.95&\(9.801 \frac{\text{T}}{\text{m}} \cdot \frac{q}{p_0}\)&\(68.07 \frac{\text{T}}{\text{m}} \cdot \frac{q}{p_0}\)&\(4.050\text{m}\)&\(1.200\text{m}\)&\(47.10\text{m}\)&\(7.500\text{m}\)&\(200\)\\
  \end{tabular}
  \end{ruledtabular}

  \end{table*}
\section{\label{sec:results}Results}
\subsection{Noise on Time-of-Flight Signal from Synchrotron Radiation}
\begin{figure}[t]
  \includegraphics[width = 0.5\textwidth]{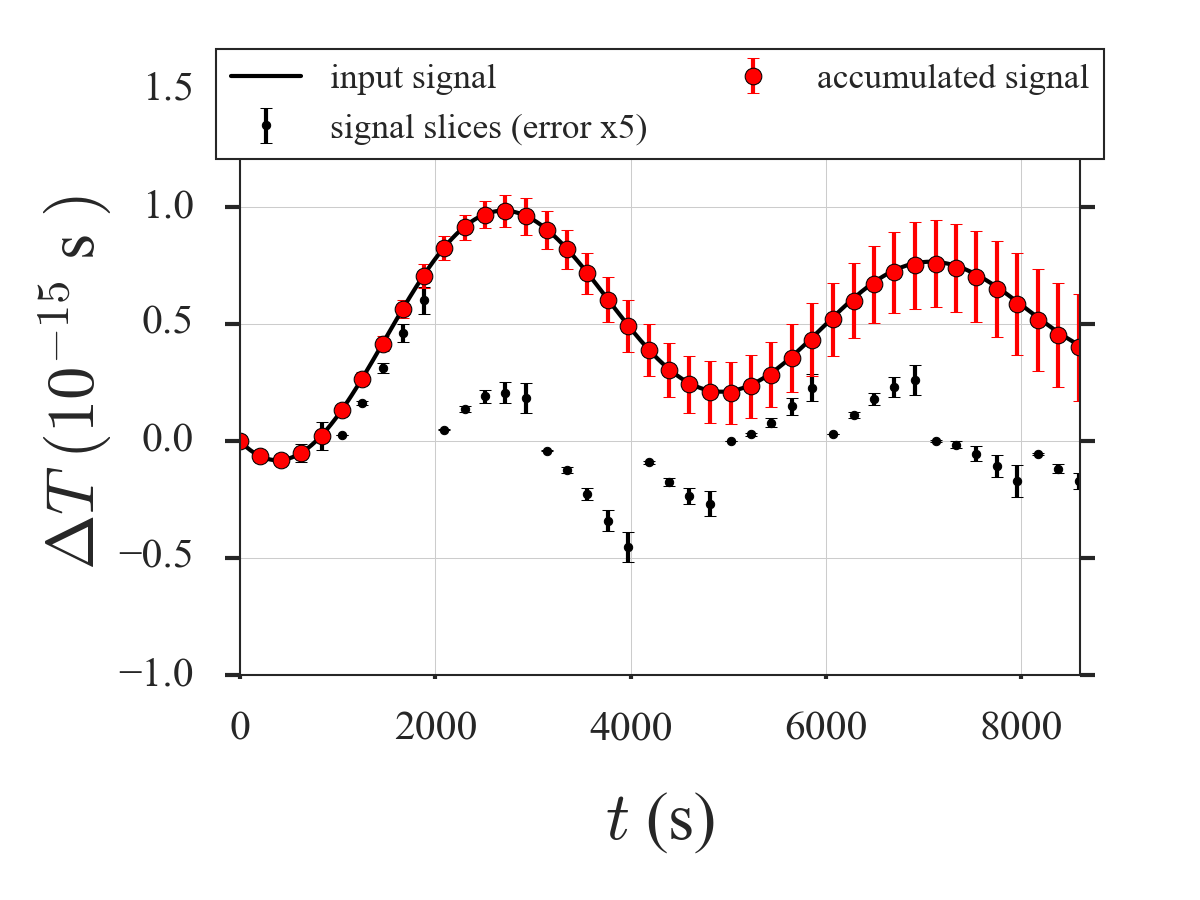}
  \caption{\label{fig:signal_error}Average arrival time expected for a perfect ring (black line) and for a simulation with synchrotron radiation (black dots) using 100 \(\text{U}_2^{+/-}\) ions at \(\beta=0.32\) and the error bars denote the error of the mean (black, scaled by factor 5 for better visibility). After \({\Delta t = 10^3\text{ s}}\) the accumulated time deviation of the ions is reset and the computation continued. The red dots denote the accumulated signal, which traces the input signal.}
\end{figure}

Similar to the experimental setup shown in Fig.\ \ref{fig:exp_setup}, a computation is performed where \(n = 100\) ions are tracked in a storage ring, which is subject to an example GW with strain as depicted in Fig.\ \ref{fig:signal_fit}. During the time interval \(\Delta t = 10^3\) s, the particles accumulate a deviation of their revolution time both due to the force from the GW strain and the synchrotron radiation. All particles experience the same average quadratic time deviation originating from the mean synchrotron radiation power, but individual fluctuations remain, leading to an uncertainty of the mean value. Every second the average revolution time of the particles is taken as a measurement point and after the time interval \(\Delta t\), the accumulated time deviation of all particles is set to zero and the computation is continued. This corresponds to the extraction and re-sinsertion of these \(n\) particles and leads to different slices of length \(\Delta t\) of the GW strain being imprinted on the arrival times of these ions.
\\
An example of the resulting average arrival time is shown in Fig.\ \ref{fig:signal_error} for \(\text{U}_2^{+/-}\) and \(\beta=0.32\) and the magnetic unit cell dimensions of the corresponding storage rings are listed in Tab.\ \ref{tab:table1}. The mean arrival time of the \(100\) ions is shown as black dots with error bars (scaled by a factor \(5\) for better visibility), which is computed every second, but not all points are shown. Using the last measurement point of an interval \(\Delta t\) as an offset, and adding it to the computed results for the next interval, the input GW strain (black line) can be reconstructed. The result is shown in red as integrated signal including error bars, which denote an accumulated uncertainty of the time deviation, because the uncertainty of the mean value at the end of a GW strain slice contributes to an uncertainty of the mean values for the following slices in the reconstructed signal.

We stress that the results presented here only include the noise on the arrival time that is due to the synchrotron shot noise, and that we otherwise assume a perfect measurement to actually determine the arrival time of each ion. The detection method of the arrival time will constitute another noise source, which, however, is likely less related to the storage ring setup than the emitted synchrotron radiation. The results obtained in our design study indicate that the mean of the particle arrival time is well-suited to reconstruct the GW input signal, confirmed by the small error bars. The findings suggest that the particle motion for this parameter choice is dominated by the GW strain rather than the noise imposed by the synchrotron radiation and it should in principle facilitate the measurement of GWs with ion chains circulating in a storage ring.

\subsection{Noise Power Spectrum for Measurement of Gravitational Waves}
\begin{figure}[t]
  \includegraphics[width = 0.5\textwidth]{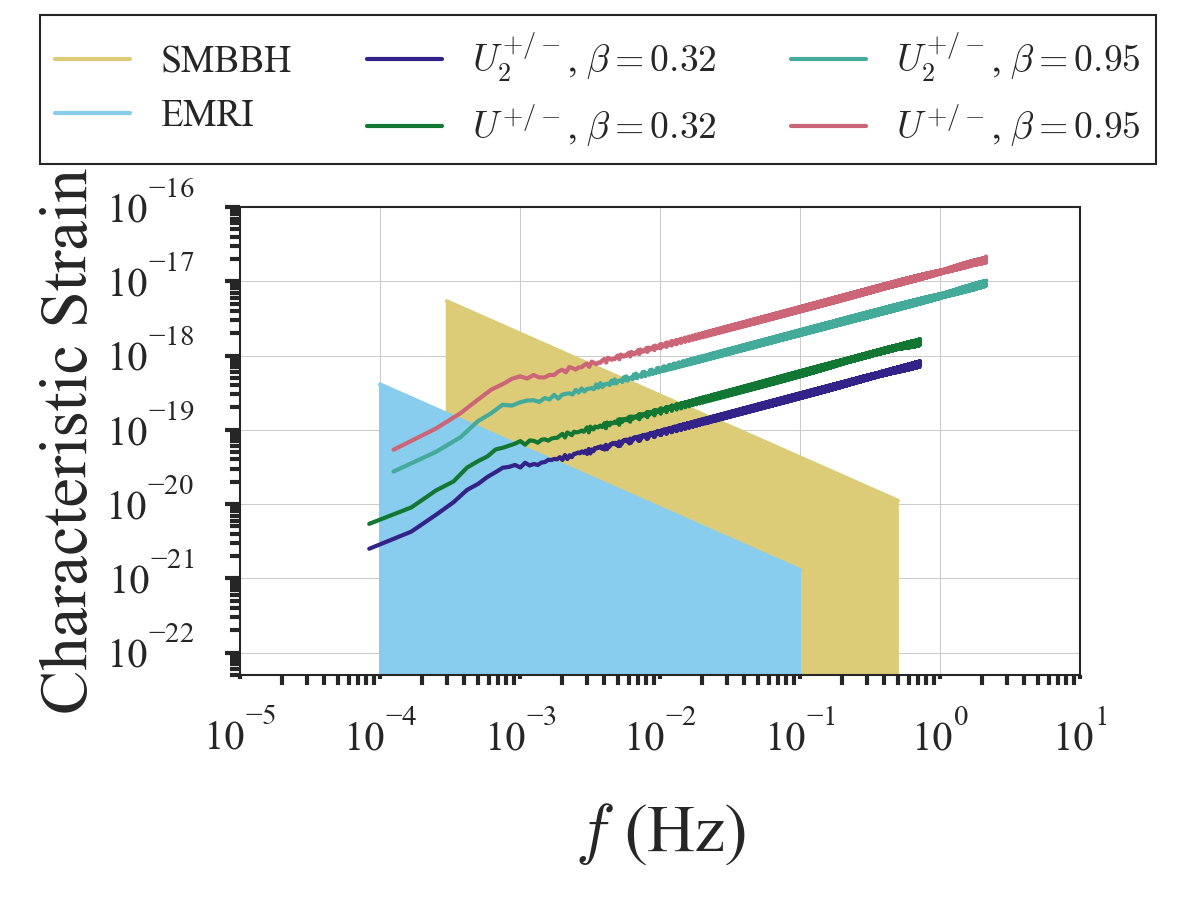}
  \caption{\label{fig:noise_sensitivity_all} Results of the noise amplitude computed from the power spectral density obtained from noise simulations for \(\text{U}_2^{+/-}\) and \(\text{U}^{+/-}\) at velocities \(\beta=0.32\) and \(\beta=0.95\). The characteristic source strain for SMBBH and EMRI obtained from \cite{Moore2015} is found to be larger in the frequency range \(10^{-4}-10^{-2}\) Hz than the noise amplitude expected from the shot noise due to synchrotron radiation.}
\end{figure}

The noise curve due to synchrotron radiation of the arrival time of \(n=100\) particles is computed by using the method described above for a duration of \(9000\) s for \(\beta=0.95\) and for a duration of \(12000\) s for \(\beta=0.32\), but without a force from a GW, using the storage rings listed in Tab.\ \ref{tab:table1}. One time step is taken as the duration of \(5000\) revolutions for all cases, which is shorter for higher velocities (\(\beta=0.95\)), such that the particles have to be tracked for more revolutions to cover the same time interval. For the cases with higher velocities, more emitted photons have to be taken into account, adding to the complexity of the numerical computation. This ultimately limits the length of the time interval that can be computed and thus, the frequency spectrum at the low end, but allows for the computation of higher frequencies. For the slower particles (\(\beta=0.32\)), however, the longer time step of the computation limits the spectrum from above. The accumulated timing deviation is set to zero every \(\Delta t = 10^3\) s to account for dumping and re-insertion of the particles. The particles exhibit the effective quadratic timing deviation from the average synchrotron radiation emission, which is subtracted from the individual timing deviation. The mean of the particle arrival time then tends to zero, since no force acts on the ions collectively.

As is discussed in Appendix \ref{app:power_spectrum}, the noise \(n(t) \sim \frac{\text{d}}{\text{d}t}\Delta T(t)\) and the one-sided power spectral density of the noise \(S_n(f) = \frac{2}{T}|\tilde{n}(f)|^2\), where \(T\) is the time interval over which the noise is recorded and \(\tilde{n}(f)\) is the Fourier transform of \(n(t)\). The noise amplitude is then given as \(h_n(f) = \sqrt{\frac{f \; S_n(f)}{\mathfrak{R}}}\) with the sky and polarization averaged sensitivity \(\mathfrak{R}=\frac{4}{15}\) \cite{Moore2015}, which is derived in the Appendix \ref{appsup:toy_model_power_spectrum} for the geometry of a storage ring. Numerically, the timing deviation of each of the \(100\) particles is simulated for each time step, from which the deriative is computed. The Fourier transformation of the result is computed, leading to the noise amplitude shown in Fig.\ \ref{fig:noise_sensitivity_all}, as averaged over all ions. The characteristic source strain for super-massive black hole binaries (SMBBH) and extreme mass-ration inspirals (EMRI) are also shown, as estimated by a suitable model \cite{Moore2015}.
\\
In general, the noise of the particle arrival time is qualitatively very similar for all cases, but slower and heavier particles show less noise than faster and lighter ones. It is found that for the frequencies below \(10^{-2}\) Hz, the characteristic strain of the sources is predicted to surpass the noise due to the synchrotron radiation, even for the faster and lighter ions. From the approximation of the analytical results, Eq.\ \ref{psd_second_approximation}, it is found that the noise strain has the proportionality \(\sim f^{3/2}\) for \(f<1/\Delta t\) and  \(\sim f^{1/2}\) for \(f>1/\Delta t\), which explains the frequency dependence of the noise and the change observable in Fig.\ \ref{fig:noise_sensitivity_all}, where \(f \sim 1/\Delta t = 10^{-3}\) Hz. In conclusion, the expected noise due to synchrotron radiation imprinted on the measurement of the characteristic strain of a GW via the average arrival time of \(n =100\) \(\text{U}^{+/-}\) or \(\text{U}_2^{+/-}\) ions should allow the detection of the strongest SMBBH or EMRI signals in principle, using a storage ring with a size comparable to that of the LHC.

\section{\label{sec:summary}Summary and Conclusion}
This work is a first quantification of the role of GW strain on the time-of-flight signal of a particle in a {\em real} storage ring. We have identified the emission of synchrotron radiation as a fundamental noise that limits the signal-to-noise ratio of a mHz GW detector based on this principle. We therefore propose an experiment based on a chain of singly-charged uranium ions or diuranium molecular ions circulating in a storage ring with a circumference similar to that of the LHC and with velocities \(\beta c\) of \(\beta =0.32\) and \(\beta=0.95\). This storage ring should be operated without energy restoration by a radio-frequency cavity, such that the free fall of the particles in the longitudinal direction is ensured. For such a setup we predict a frequency window of \(10^{-4}-10^{-2}\) Hz, in which a measurable signal of the GW strain from SMBBH and EMRI type sources in the storage ring could be expected despite the synchrotron emission noise of the ions.

In order to facilitate the numerical computations, we have re-derived the equations of motion used for particle tracking for a particle passing through the different magnetic elements of a storage ring with a FODO magnetic lattice under the influence of an external force and for a particle emitting a number of photons during the passage of a sector magnet. We describe a numerical method to compute the number and energy of the photons emitted per turn via synchrotron radiation using the classical energy spectrum and average emitted power. The power spectral density of the synchrotron radiation noise is discussed in detail -- analytically by the definition of appropriate models and numerically as the result of simulations. The findings of this work give a first estimation of which parameters are required to have a chance of measuring GWs with the time-of-flight signal of circulating particles in a storage ring, and of the limitations of the detector performance due to its operating principle. These results, however, are of course not enough to build a functioning detector, be it by completely designing and building a new storage ring from the ground up or by re-purposing an existing one. 

First, the detection principle of the ion arrival time, as they circulate around the ring, is yet to be determined, which will be part of a future work and which has to be quantified in order to obtain a more realistic estimate of the theoretically expected signal-to-noise ratio. Second, should a measurement technique with the required temporal or spatial resolution be found, it will certainly put additional constraints on the particle type and velocity, such that the performance estimates may change. Third, as another important noise source, the gravitational gradient noise has to be quantified, which is a central and likely a dominant noise source in the mHz frequency range, as it is for any other Earth-based detector design. Finally, the operation according to the proposed experimental setup requires the generation, injection and circulation of the particles to exhibit a very small deviation from the design energy. Together with other noise sources present in an actual storage ring, this further limits the performance of the proposed detector design and its experimental realisation is still far from clear. Nevertheless, the outlook for detecting mHz GWs with a storage ring-based detector is still promising, and we regard this work -- with its focus on the reduction of the synchrotron emission impact on the arrival time measurements -- as a first step towards a concrete design study. This work also provides valuable input to future studies that will address the generation and acceleration of the ions, as well as their arrival time measurement.

\begin{acknowledgments}
TS would like to thank Mikhail Korobko, Reinhard Brinkmann, Danyal Winters and Andreas Wolf for helpful discussions. The authors acknowledge support by the Deutsche Forschungsgemeinschaft (DFG, German Research Foundation) under Germany’s Excellence Strategy – EXC 2121 `Quantum Universe' – 390833306.
\end{acknowledgments}

\appendix
\section{Relating Signal in Storage Ring to Ring with Fixed Radius}
\subsection{\label{app:relating_signals}Analytic Results}
A key assumption of the idealized model discussed by Rao et al. \cite{Rao2020} is that the radius of the trajectory of the traveling particle is constant while it is subject to the GW strain. In a storage ring, this assumption is invalid, as the de- or increased momentum of a particle due to an external force also leads to a change of the relativistic mass \(m \gamma\), causing a change in the radius of curved paths in magnetic fields and thus a non-linear change of the circulation time. The change of circulation frequency \(\Delta \omega\) for a nominal circulation frequency \(\omega_0\) under momentum change is given by \(\frac{\Delta \omega}{\omega_0} =\eta \frac{\Delta p}{p_0}\), where \(p_0\) is the nominal momentum, \(\Delta p\) the momentum deviation, the slip factor is denoted by \(\eta := \frac{1}{\gamma^2} - \alpha_p\) and the momentum compaction factor by \(\alpha_p\) \cite{Hinterberger2008}. Thus, the deviation \(\Delta t\) from the nominal revolution time after circulating once around the storage ring is given as
\begin{align}
\Delta t &= - \eta \gamma^2 T_0 \frac{\Delta v}{v_0} = \eta \gamma^2 \frac{\Delta L}{v_0}.
\label{dt_ring}
\end{align}
In the second equality, a shorter round trip time is related to an advance \(\Delta L>0\) of a particle relative to a design particle, leading to a time deviation \(\Delta t>0\) between the two, removing the sign. The time dependence is not written explicitly. Next, the acceleration is approximated as
\begin{align}
a_\parallel(t) & \approx -\dot{h}_{\theta\phi\psi}(t)v_0, 
\end{align}
because \(h_{\theta\phi\psi}(t) \ll 1\). In a ring with fixed radius the time integral of the longitudinal acceleration \(a_\parallel\) directly leads to \(\Delta v(t)\), which, integrated over time again and divided by the nominal velocity \(v_0\), results in the timing deviation \(\Delta t\). Thus, one finds
\begin{align}
\Delta v(t) &= -v_0\int_0^{t} \dot{h}_{\theta\phi\psi}(t')\; dt'
\label{velocity_fixed}
\end{align}
and therefore
\begin{align}
\Delta L(t) &= -v_0\int_0^t\left( h_{\theta\phi\psi}(t') - h_{\theta\phi\psi}(0)\right)dt',\label{runtime_deviation}
\end{align}
which, together with Eq.\ \ref{dt_ring}, leads to
\begin{align}
\Delta T^\text{measure}(t) & = -\eta \gamma^2\int_0^t\left( h_{\theta\phi\psi}(t') - h_{\theta\phi\psi}(0)\right)dt'. \label{equality_fixed_ring}
\end{align}
The predicted signal for a fixed ring is given by \cite{Rao2020}
\begin{align}
\Delta T^\text{fixed}(t) &= \left(1-\frac{v_0^2}{2c^2}\right) \int_0^t\left(h_{\theta\phi\psi}(t,\alpha(t))\nonumber\right.\\
&\left.-h_{\theta\phi\psi}(t_0,\alpha_0)\right)dt.
\label{suvrat_signal}
\end{align}
By comparing Eq.\ \ref{suvrat_signal} to Eq.\ \ref{equality_fixed_ring}, it follows
\begin{align}
\frac{\Delta T^\text{fixed}(t)}{1-\frac{v_0^2}{2c^2}} & = -\frac{1}{\eta}\frac{1}{\gamma^2}\Delta T^\text{measure}(t).
\end{align}
In an experiment, \(\Delta T^\text{measure}\) can only be detected for changing space time and not for a constant one. In the case of a computation, where the signal \(\Delta L^\text{Ring}\) is obtained in relation to the design particle\footnote{In this section, the superscript \(^\text{Ring}\) denotes that a quantity is given relative to a design particle, as it is common in transfer matrix formalism.}, however, even for a constant space time, a linearly increasing result is computed, because the space time for the design particle is always flat and thus the initial space time strain has to be accounted for in the transformation. From Eq.\ \ref{velocity_fixed} follows,
\begin{align}
\Delta v^\text{Ring}(t) &= - v_0 \int_0^t \dot{h}_{\theta\phi\psi}(t')\;dt' + \Delta v(0),
\end{align}
where the constant
\begin{align}
\Delta v(0) &= -v_0 \overline{h_{\theta\phi\psi}(0)},
\end{align}
which is given by the average space time strain at \(t=0\) during the time \(T_0\), originates from the comparison of the particle in the ring to the design particle. This leads to
\begin{align}
  \Delta L^\text{Ring}(t) &= - v_0 \int_0^th_{\theta\phi\psi}(t')dt'.
\end{align}
Using this result, one can relate the result of the computation for a storage ring \(\Delta T(t)^\text{Ring}\) to the prediction signal for fixed radius \(\Delta T(t)^\text{fixed}\) via
\begin{align}
\Delta T^\text{fixed}(t)=  &- \frac{1}{\eta}\frac{1}{\gamma^2}\left(1 - \frac{v_0^2}{2c^2}\right) \Delta T^\text{Ring}(t) \nonumber \\
&- \left(1 - \frac{v_0^2}{2c^2}\right) \overline{h_{\theta\phi\psi}(t_0,\alpha_0)} t, \label{relation_signal_strength}
\end{align}
where the time dependence has been writen explicitly and \(h_{\theta\phi\psi}(0)\) has been replaced with the average \(\overline{h_{\theta\phi\psi}(t_0,\alpha_0)} = \frac{1}{N}\sum_{i\in\text{Ring}}^Nh_{\theta\phi\psi}(t_0,\alpha_i)\) over the positions \(\alpha_i\) for all \(N\) ring elements (magnets and drift spaces).

\subsection{\label{app:relating_signals_numerics}Numerical Results}
\begin{figure}[ht]
  \includegraphics[width = 0.5\textwidth]{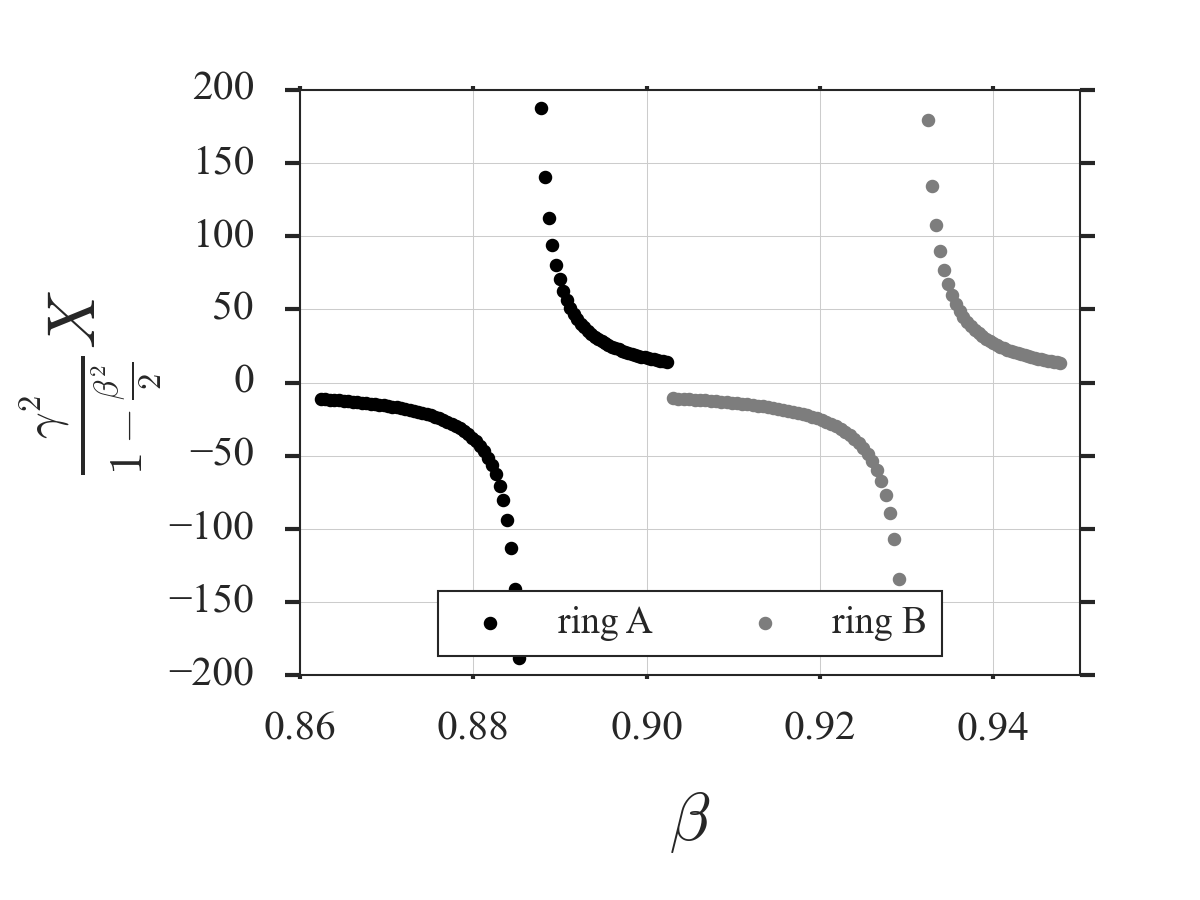}
  \caption{\label{fig:relative_factor} The unit cell design of a storage ring directly relates \(\beta\) of the particle and the slip factor \(\eta\), which leads to the signal strength. Both ring A and ring B have a circumference of \({120\text{ m}}\). Ring A has a transition energy of \(\beta^\text{transition} \approx 0.886\) and ring B of \(\beta^\text{transition} \approx 0.931\), where the signal vanishes and the scaling factor diverges as \(\eta \rightarrow 0\).}
\end{figure}

\begin{table*}
  \caption{Parameters for the two proton rings A and B with \(120\) m circumference and 12 unit cells. The magnetic unit cells consist of transversally focusing (QF) and defocusing (QD) quadrupole magnets with magnetic parameters \(k^{\text{QD}/\text{QF}}\), sector magnets with edge focusing and drift spaces, each with lengths \(l^\text{sector}\) etc.\label{tab:table2}}
  \begin{ruledtabular}
  \begin{tabular}{cccccccc}
   &\(k^\text{QF}\)&\(k^\text{QD}\)&\(l^\text{QD}\)&\(l^\text{QF}\)&\(l^\text{sector}\)&\(l^\text{drift}\)&\(\beta^\text{transition}\)\\ 
   \hline
   ring A&\(1.782 \frac{\text{T}}{\text{m}} \cdot \frac{q}{p_0}\)&\(5.768 \frac{\text{T}}{\text{m}}  \cdot \frac{q}{p_0}\)&\(0.380\text{m}\)&\(0.228\text{m}\)&\(2.250\text{m}\)&\(1.128\text{m}\)&\(0.886\)\\
   ring B&\(3.119 \frac{\text{T}}{\text{m}} \cdot \frac{q}{p_0}\)&\(2.884 \frac{\text{T}}{\text{m}} \cdot \frac{q}{p_0}\)&\(0.344\text{m}\)&\(0.588\text{m}\)&\(2.250\text{m}\)&\(1.056\text{m}\)&\(0.931\)\\
  \end{tabular}
  \end{ruledtabular}
  \end{table*}
The result for \(\Delta T^\text{fixed}\) is compared to the result from the storage ring \(\Delta T(t)^\text{Ring}\) and the transformation Eq.\ \ref{relation_signal_strength} is verified. Two different FODO configurations, suitable for protons, are used, allowing for smaller ring sizes and lower kinetic energies, which we choose as a simple case to study the transformation. 
\begin{figure}
  \centering
    \includegraphics[width=\linewidth]{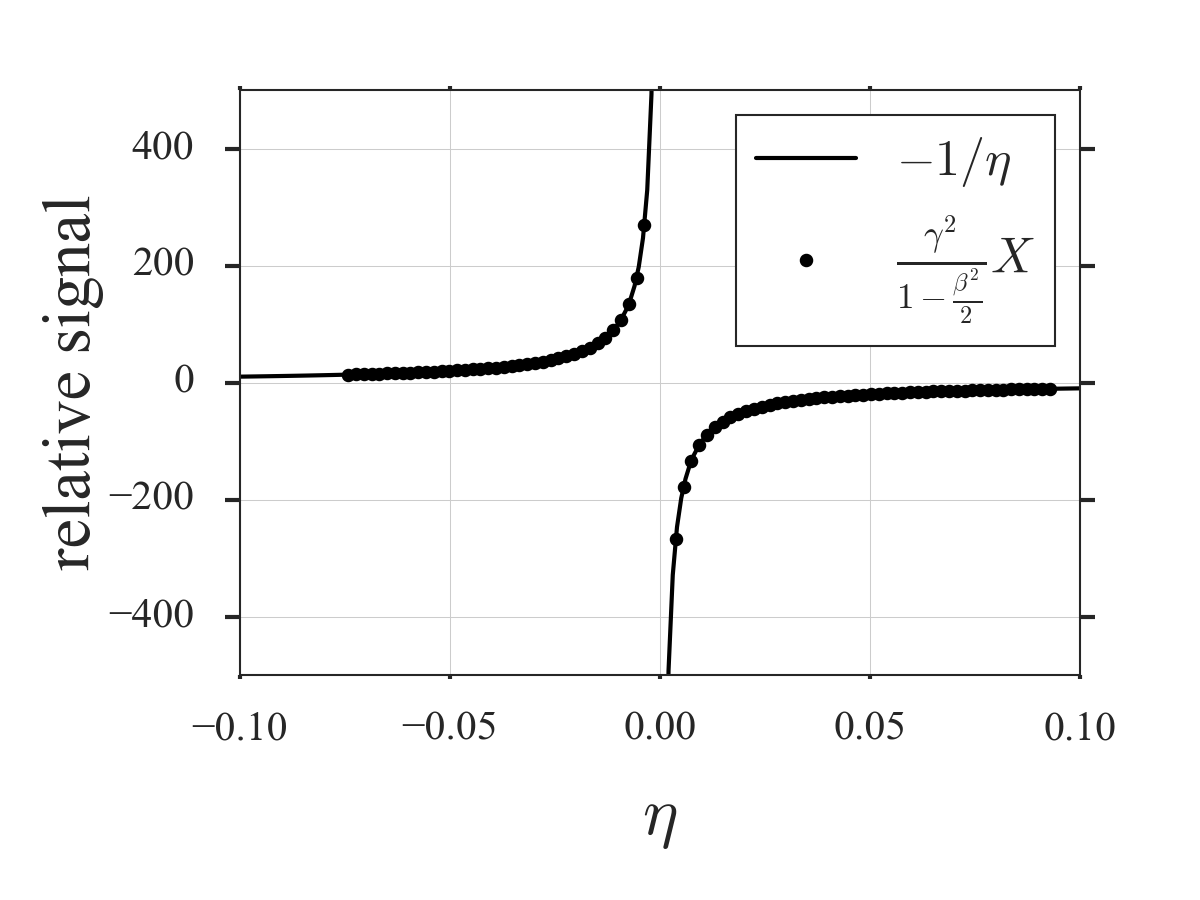}
  \caption{\label{fig:factor_1}The parameter \(X\) used in the fit for the relative signal strength in Eq.\ \ref{fitting_function}, as the slip factor \(\eta\) is varied, and compared to \(-1/\eta\)}.
\end{figure}
\begin{figure}
  \centering
    \includegraphics[width=\linewidth]{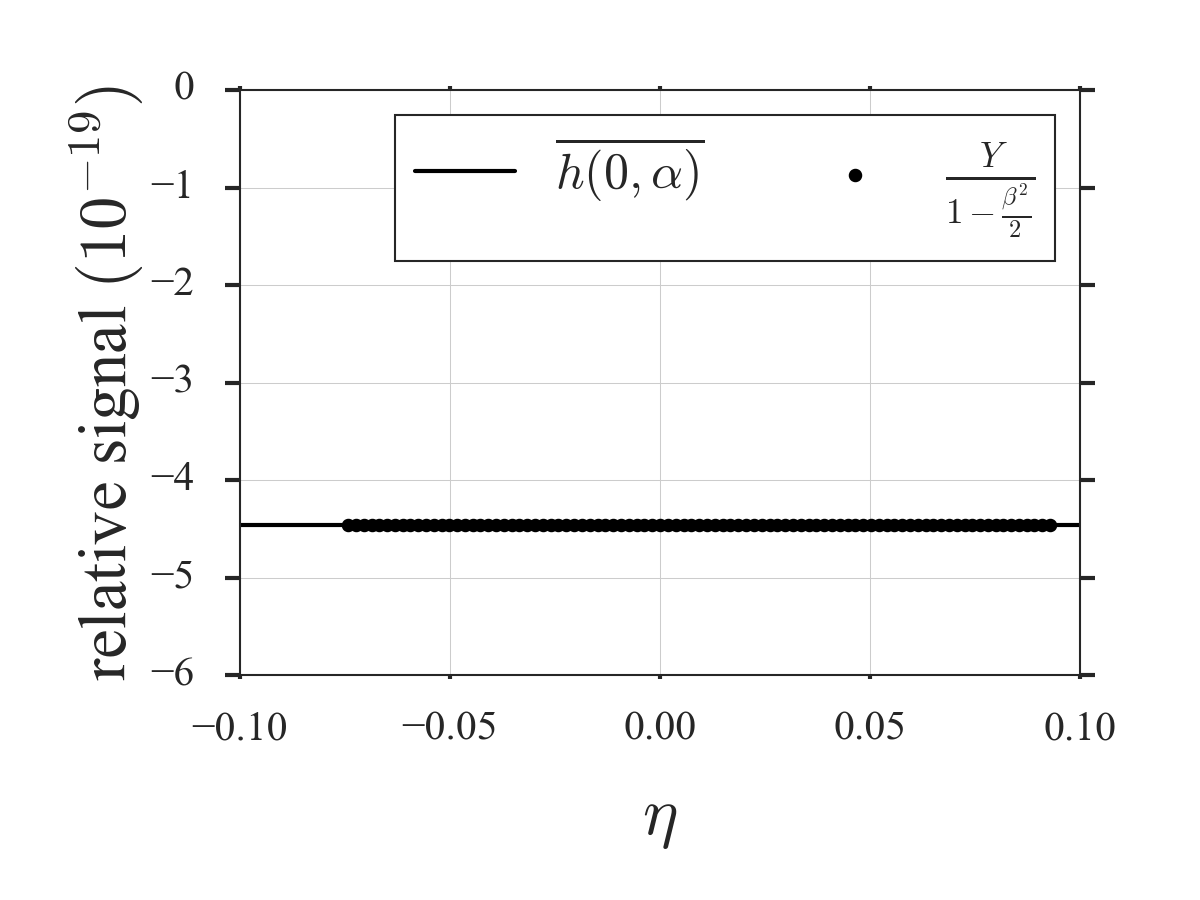}
  \caption{\label{fig:factor_2}The parameter \(Y\), used in Eq.\ \ref{fitting_function}, as the slip factor \(\eta\) is varied. The constant value for \(\frac{Y}{1-\frac{1}{2}\beta^2}\) is equal to the average GW strain around the ring shown in Fig.\ \ref{fig:strain_around_ring}, which is independent of the ring geometry and particle velocity.}
\end{figure}
Thus, two model rings A and B for protons are designed, each consists of 12 unit cells with \({10 \text{ m}}\) length and the parameters for the FODO cells listed in Tab.\ \ref{tab:table2}. Ring A has a transition point at \(\beta^\text{transition} \approx 0.886\) and ring B close to \(\beta^\text{transition} \approx 0.931\). Particles with velocities ranging from \(\beta = 0.903\) to \(\beta= 0.948\) are all computed for ring B, such that some velocities operate close to the transition point, where \(\eta = 0\). For all cases the GW signal and the resulting acceleration are identical and therefore only the impact on the timing deviation changes. The result for each simulation of a storage ring are fitted to the predicted signal of a perfect ring using 
\begin{equation}
  \Delta T^\text{fixed} = X \; \Delta T(t)^\text{Ring} + Y \; t,
  \label{fitting_function}
\end{equation}
resulting in Figs. \ref{fig:factor_1} and \ref{fig:factor_2} for ring B. As shown by the fit, the two parameters \(X\) and \(Y\) depend on \(\beta\) and \(\gamma\), such that dividing them by their corresponding factors directly leads to \(\frac{\gamma^2}{1-\frac{1}{2}\beta^2}X =-\frac{1}{\eta}\) and \(\frac{1}{1-\frac{1}{2}\beta^2}Y= \overline{h(0,\alpha)}\), consistent with Eq.\ \ref{relation_signal_strength}. The timing deviation signal due to the GW force is expected to vanish, when the ring is operated close to the transition point, such that the factor required to scale up the signal to that of the perfect ring diverges, which is also confirmed by the fit. The results for ring A and B plotted against \(\beta\) are shown in Fig.\ \ref{fig:relative_factor}. Due to the different ring geometry, the transition point of ring A is different than that of ring B. Therefore, the signal strength at a specific \(\beta\) can be tuned by the ring geometry. It is important to ensure that particles at sufficiently low velocities are operated in storage rings away from the transition energy, to show a non-vanishing signal.

\section{Transfer Matrices with External Force}
\label{app:matrices}
To analyse the effect of the GW strain on a particle circulating in a storage ring, the standard textbook procedure \cite{Hinterberger2008} for the matrix formalism of first-order equations of motions is modified to include a constant external force, which will be derived in this section. The main result will be the expressions for the position and
momentum shifts, which will have to be applied on the phase space vector after a transfer matrix, to correctly account for the force of the GW in a storage ring element. The state vector of such a transformation is given by the coordinates \(\{x,x',y,y',l,\delta\}\), where the first two components are the radial position offset and radial velocity, the third and fourth component are the axial position offset and velocity, and the remaining two are the longitudinal offset and momentum deviation from a design particle with momentum \(p_0\). \\
\subsection{\label{app:transversal_eom}Transversal Equation of Motion for Sector Magnet}
The storage ring is assumed to be in the \(x,z\)-plane, where the two transversal directions are \(x\) and \(y\). The transversal motion of a particle in a magnetic field is given by \begin{align}
  \ddot{x} - \omega^2 (\rho_0 + x) & = - \frac{q}{p} v_s^2 B_y \label{x_eq} \\
  \ddot{y} & = + \frac{q}{p} v_s^2 B_x. \label{y_eq}
\end{align}
The longitudinal velocity in the direction of motion of the particle is given by 
\begin{equation}
  v_s = \left(\rho_0 + x\right)\omega + A s, \label{longitudinal_velocity}
\end{equation}
where the additional term
\begin{equation}
A := \frac{\langle F \rangle}{m \gamma ^3}\frac{1}{\beta c} \label{a_expression}
\end{equation}
(in units \(\frac{1}{\text{s}}\)) denotes the acceleration, due to a force \(\langle F\rangle\), as the particle progresses on its path, parametrized by the path length \(s\), which is obtained, via the regular laws from classical mechanics and where \(\rho_0\) is the radius of the trajectory and \({\omega = v_0/\rho_0}\) the angular velocity. The resulting equation of motion in linear order for \(x\) is then
\begin{align}
x'' + \frac{1-n}{\rho_0^2}x &=\frac{\delta}{\rho_0}-\frac{2 A z}{\rho_0^2 \omega}. \label{x_eq_result}
\end{align}
Eq.\ \ref{x_eq_result} is basically the standard result, with an additional constant term on the right hand side and for the \(y\)-component the same result as that of the unperturbed particle in a ring is found. The Green's function approach is employed for the full solution
\begin{equation}
x(s) = x_0 c_x(s) + x_0's_x(s) + \delta d_{x1}(s) + d_{x2}(s), 
\end{equation}
where a new term \(d_{x2}(s)\) compared to the usual result \cite{Hinterberger2008} occurs, that depends on the external force and magnet geometry, but not on the coordinates of the particle, given by
\begin{align}
d_{x2}(s)&:= - \frac{2A}{\rho_0^2\omega}\left[\frac{s}{k} - \frac{\sin{\sqrt{k}s}}{k^2}\sqrt{k}\right] \label{eq_dx2}
\end{align}
with the derivative
\begin{align}
d_{x2}'(s) &:= - \frac{2A}{\rho_0^2\omega}\left[\frac{1}{k} - \frac{\cos{\sqrt{k}s}}{k}\right]. \label{eq_dx2_d}
\end{align}
Here \(k = 1/\rho_0^2\). Since this term does not depend on any of the phase space coordinates of the state vector it cannot be represented by a matrix element in the transfer matrix of the sector magnet. 

\subsection{\label{app:eom}Longitudinal Equation of Motion for Sector Magnet}

The longitudinal equation of motion for a particle traveling from \(0\) to \(s\) in a circular arc is given by \cite{Hinterberger2008}
\begin{align}
l(s) - l(0) & = -(S-s) + s \frac{\Delta v}{v_0} + \Delta l \label{longitudinal_eq_o_m}.
\end{align}
The longitudinal displacement \(\Delta l\) of a free particle, subject to a constant acceleration \(a\) during the time interval \(\Delta t\), which it requires to cover the distance \(s\), is given by \(\Delta l = \frac{1}{2}a \Delta t^2\), which can be rewritten as
\begin{align}
\Delta l & = \frac{\langle F \rangle}{2 m \gamma^3 c^2}\left(\frac{s}{\beta}\right)^2\label{eq_longitudinal_offset}
\end{align}
accounting for the slow-down of the particle due to the external force. Using \(\frac{\Delta v}{v_0} = \frac{1}{\gamma^2} \frac{\Delta p}{p_0}\), Eq.\ \ref{longitudinal_eq_o_m} is rewritten as
\begin{align}
l(s) - l(0) &= - \int_0^s \left(h(\bar{s})\; x(\bar{s})\;d\bar{s}\right) + \frac{s}{\gamma^2}\delta + \Delta l. \label{longitudinal_eq_o_m_2}
\end{align}
Again, the usual matrix elements for the transfer matrix are derived, with an additional term
\begin{align}
R(s) &= \frac{\langle F \rangle}{2 c^2 \beta^2 m \gamma^3}s^2 \nonumber \\
&+ \frac{2 \langle F \rangle}{\rho_0^2 \beta^2 c^2 m\gamma^3}\left[\frac{s^2}{2k} +\left(\frac{\cos{\sqrt{k}s}}{k^2} - \frac{1}{k^2}\right)\right], \label{matr_elem_4}
\end{align}
which does not depend on the coordinates of the phase space vector and has to be added together with the offset from Eqs. \ref{eq_dx2} and \ref{eq_dx2_d} after matrix multiplication to correctly account for the position and velocity of the particle after passage of the sector magnet.
\subsection{\label{app:subsec}Longitudinal Equation of Motion for Drift Space}
For a free particle in a drift space, the shift \(\Delta l (s)\) from Eq.\ \ref{eq_longitudinal_offset} occurs, leading to the equations of motion
\begin{align}
l(s) &= l_0 + \frac{s}{\gamma^2}\delta_0 + \Delta l(s) \label{eq_longitudinal_drift_modified}\\
\delta(s) &= \delta_0 + \Delta \delta(s) \label{eq_momentum_drift_modified}
\end{align}
with the momentum deviation \(\delta\), where the external force leads to a shift
\begin{align}
\Delta \delta (s) = \Delta \left(\frac{\Delta p}{p_0}\right) = \frac{\langle F \rangle}{p_0 c}\frac{s}{\beta}. \label{eq_momentum_shift}
\end{align}
For the quadrupole magnets this result is identical, because only the transversal equations of motion differ from those of a drift space.
\subsection{Transfer Matrices and Shifts}
Summarising, the shifts due to the external force are stated explicitly.
\subsubsection{Sector Magnet}
The transformation is given by the regular transfer matrix \(R^\text{sec}(l)\) and the addition of a shift, resulting in
\begin{align}
x(l) & = R^\text{sec}(l) x(0) + \Delta x^\text{sec}(l),
\label{sectormagnet_equation}
\end{align}
where \(l\) is the length of the sector magnet and the  shift is given by
\begin{align}
&\Delta x^\text{sec}(l) \nonumber\\
&=\begin{pmatrix}
  - \frac{2}{\rho_0 \beta c}\frac{\langle F \rangle}{m \gamma^3 \beta c}\left[\frac{l}{k} - \frac{\sin{\sqrt{k}l}}{k^2}\sqrt{k}\right] \\
  - \frac{2}{\rho_0 \beta c}\frac{\langle F \rangle}{m \gamma^3 \beta c}\left[\frac{1}{k} - \frac{\cos{\sqrt{k}l}}{k}\right]\\
  0\\
  0\\
  \frac{\langle F \rangle}{2 m \gamma^3 \beta^2 c^2}l^2 + \frac{2}{\rho_0^2 c^2 \beta^2}\frac{\langle F \rangle}{m \gamma^3}\left[\frac{l^2}{2k} +\left(\frac{\cos{\sqrt{k}l}}{k^2} - \frac{1}{k^2}\right)\right]\\
  \frac{\langle F \rangle}{p_0 c}\frac{l}{\beta} 
\end{pmatrix}.
\label{displacement_sector}
\end{align}
\subsubsection{Drift Space and Quadrupole Magnet}
The transformation for the drift space and focusing (QF) and defocusing (QD) quadrupole magnets are given using their respective transfer matrix \(R(l)\) as
\begin{align}
  x(l) & = R^\text{drift}(l) x(0) + \Delta x^\text{drift}(l)
\end{align}
and
\begin{align}
  x(l) & = R^\text{QD/QF}(l) x(0) + \Delta x^\text{drift}(l),
\end{align}
but in both cases the displacement vector \(\Delta x^\text{drift}\) is given by
\begin{align}
  \Delta x^\text{drift}(l)
  =\begin{pmatrix}
   0 \\
    0\\
    0\\
    0\\
    \frac{\langle F \rangle}{2 m \gamma^3 \beta^2 c^2}l^2\\
    \frac{\langle F \rangle}{p_0 c}\frac{l}{\beta} 
  \end{pmatrix},
  \label{displacement_drift}
  \end{align}
using Eqs. \ref{matr_elem_4} and \ref{eq_momentum_shift}.

\section{\label{app:synchrad}Photon Emission}
\subsection{\label{appsup:inverse_function} Inverse Transformation for Radiation Sampling}
The procedure is based in \cite{Arvo2001} and was adapted to the regular sphere, required for the sampling of the synchrotron radiation. By parametrising the surface of a sphere via \(\vec{r}(\theta,\phi)\),
the function
\begin{align}
\sigma(\theta,\phi) &= | \frac{d \vec{r}}{d\theta} \times \frac{d \vec{r}}{d\phi}|
\end{align}
is computed, used in the definition of two cumulative distribution functions
\begin{align}
G(s) &:= \frac{\int_0^{2\pi} \int_0^s \sigma(\theta,\phi) \; d\theta \; d\phi}{ \int_0^{2\pi} \int_0^\pi \sigma(\theta,\phi)\; d\theta \; d\phi}\\
H(\theta,t) &:= \frac{\int_0^t \sigma(\theta,\phi) \; d\phi}{ \int_0^{2\pi} \sigma(\theta,\phi) \; d\phi}.
\end{align}
These functions are next inverted, such that
\begin{align}
g(s_1) &= G^{-1}(s_1)\\
h(s_1,s_2)&=H^{-1}(s_1,s_2).
\end{align}
While this procedure is more general, in the particular case of a sphere it follows that \(H^{-1}(s_1,s_2) = \text{Id}(s_2)\). The inverted functions can be used to map a distributions of angles \(\{\theta,\phi\} \in [0,\pi] \times [0,2\pi]\) onto itself, by \((g(s_1),h(s_1,s_2)):[0,\pi] \times [0,2\pi] \rightarrow [0,\pi] \times [0,2\pi]\), which turns a stratified sampling of cartesian space into a stratified sampling of the surface of a sphere.

\subsection{\label{appsup:eom} Equations of Motion for Photon Emission in a Sector Magnet}

For determining of the equation of motion in a sector magnet under photon emission, another shift is added to the right hand side of Eq.\ \ref{sectormagnet_equation}, which encodes the shift of the phase space vector for each emitted photon with momentum \(\vec{k}=(k_x,k_y,k_z)\). In six-dimensional phase space each photon transmits the momentum
\begin{align}
\Delta_\gamma(\vec{k}) := (0,-\frac{\hbar k_x}{p_0},0,-\frac{\hbar k_y}{p_0},0,-\frac{\hbar k_z}{p_0})
\end{align}
on the particle. The transfer matrix for a sector magnet is given by a matrix product of a homogeneous bending magnet and the edge focusing matrices via \(R^\text{sec}(\alpha, \rho_0)=R^\text{edge}R^\text{hom}(\alpha, \rho_0)\left(R^{\text{edge}}\right)^{-1}\), where \(l=\alpha \rho_0\) and the edge focusing angle is \(\psi = \alpha/2\), which is not written exlicitly. Then the outgoing state vector \(x(l)\) of the particle after having passed the sector magnet and emitting a photon in dependence of the incoming state vector \(x(0)\) is computed as
\begin{align}
  x(l) & = R^\text{edge}R^\text{hom}(\alpha - \Delta \alpha, \rho_0) \times \nonumber \\
  &\left(R^\text{hom}(\Delta \alpha, \rho_0)\left(R^{\text{edge}}\right)^{-1} x(0) + \Delta_\gamma (\vec{k}) \right).
\end{align}
\begin{figure}[t]
  \includegraphics[width = 0.4\textwidth]{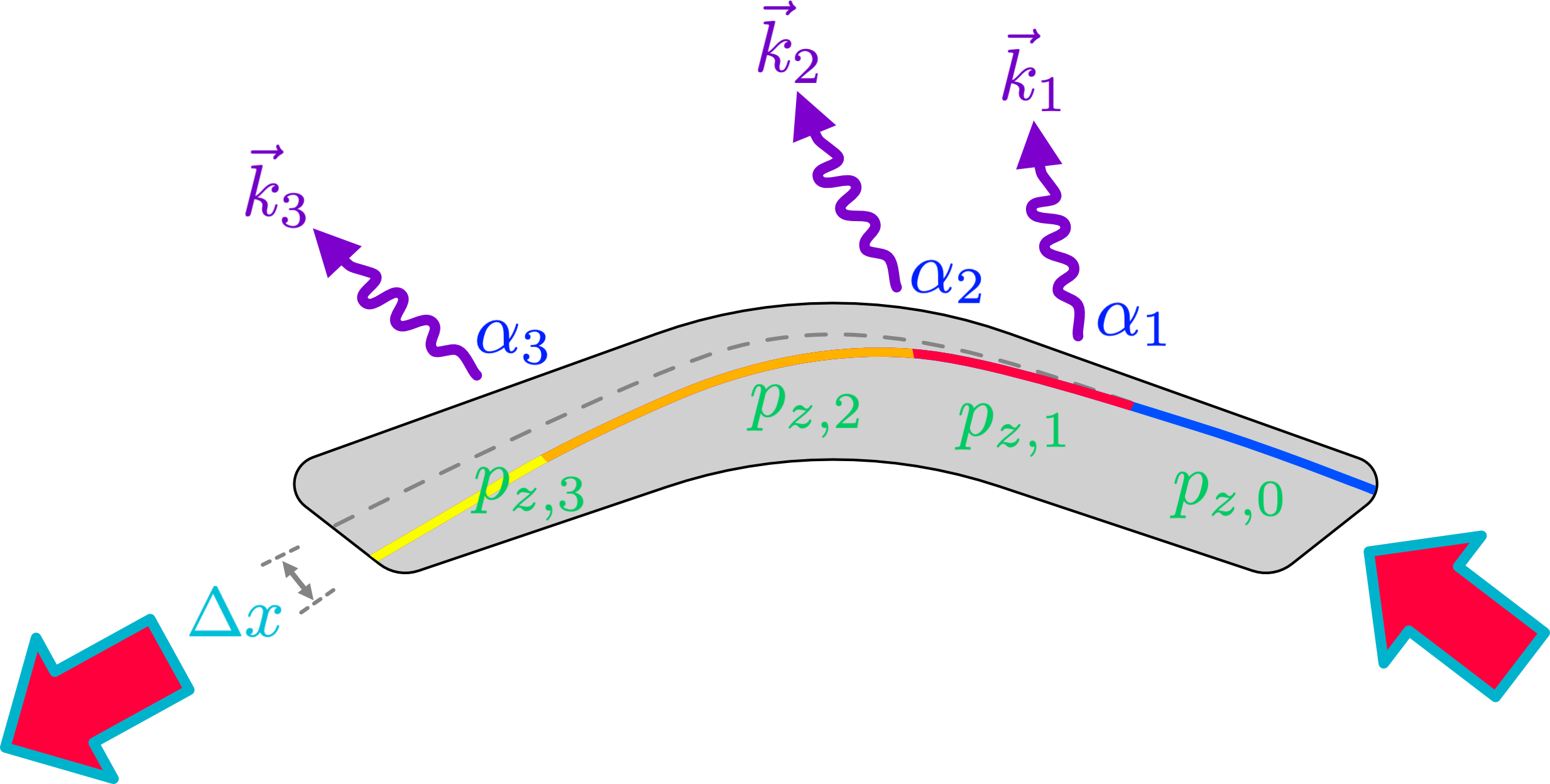}
  \caption{\label{fig:synchrotron} A particle enters a sector magnet from the right and three photons at angles \(\alpha_i\) are emitted that change the initial momentum \(p_{z,0}\) of the particle. At the end of the magnet, the radial and longitudinal position relative to a design particle change.}
\end{figure}
Here \(\Delta \alpha \in (0,\alpha)\) is the angle segment which the particle passes within the sector magnet before the photon is emitted. This procedure can be readily extended for \(n\) photon emissions. Here the particle covers the arcsegment \(\Delta \alpha_j\) in between two photon emissions, such that \(\alpha_i = \sum_{j=1}^i \Delta \alpha_j\) is the angle at which the \(i\)th photon is emitted, see Fig.\ \ref{fig:synchrotron}. The total shift \(\Delta x^\gamma (\Delta \alpha_1,\dots, \Delta \alpha_n,\vec{k}_1,\dots, \vec{k}_n)\) for \(n\) photons is then obtained via

\begin{align}
&\Delta x^\gamma (\Delta \alpha_1,\dots, \Delta \alpha_n,\vec{k}_1,\dots, \vec{k}_n) := \nonumber \\
&R^\text{edge}R^\text{hom}(\alpha - \sum_{i=1}^{n}\Delta \alpha_i, \rho_0)\left(R^\text{hom}(\Delta \alpha_n, \rho_0) \times \right. \nonumber \\
&\left( \dots \left(R^\text{hom}(\Delta \alpha_2, \rho_0) \left(R^\text{hom}(\Delta \alpha_1, \rho_0)\left(R^{\text{edge}}\right)^{-1} x(0) \right. \right.\right.\nonumber \\
&\left. \left.\left. \left.+ \Delta_\gamma (\vec{k}_1) \right) + \Delta_\gamma (\vec{k}_2) \right) + \dots \right) + \Delta_\gamma (\vec{k}_{n}) \right)\nonumber \\  
 &- R^\text{edge}R^\text{hom}(\alpha, \rho_0)\left(R^{\text{edge}}\right)^{-1} x(0),
\end{align}

which can be explicitly written as

\begin{widetext}
\begin{align}
  &\Delta x^\gamma (\Delta \alpha_1,\dots, \Delta \alpha_n,\vec{k}_1,\dots, \vec{k}_n) = \nonumber \\
  &\begin{pmatrix}
  \frac{\rho_0 \hbar}{p_0} \left(-\sum_j^n k_z^j + \sum_j^n k_z^j \cos{(\alpha - \sum_{i}^j\Delta \alpha_i)} - \sum_j^n k_x^j\sin{(\alpha-\sum_i^j \alpha_i)}\right) \\
  -\frac{\hbar}{p_0}\sec{\psi} \left(\sum_j^n k_x^j \cos{(\alpha - \psi - \sum_i^j \Delta \alpha_i)}\right) - \frac{\hbar}{p_0}\sum_j^n k_z^j \sin{(\alpha - \psi - \sum_i^j\Delta\alpha_i) -\frac{\hbar}{p_0}\sum_j^nk_z^j\tan{\psi}} \\
  -\frac{\rho_0 \hbar}{p_0}\sum_j^nk_y^j(\alpha - \sum_i^j\Delta\alpha_i)\\
-\frac{\hbar}{p_0}\sum_j^nk_y^j-\frac{\hbar}{p_0}\tan{\psi} \left( \sum_i^{n-1} \Delta\alpha_{i+1}\left(\sum_j^ik_y^j\right)+\left(\alpha - \sum_i^n\Delta\alpha_i\right)\times\left(\sum_i^n k_y^i \right)\right) \\
-\frac{\rho_0 \hbar}{p_0}\sum_j^nk_x^j\cos{(\alpha-\sum_i^j\Delta\alpha_i)} -\frac{\rho_0 \hbar}{p_0}\left(\sum_j^n k_z^j \sin{(\alpha - \sum_i^j\Delta\alpha_i)}\right) +\frac{\rho_0 \hbar}{p_0}\sum_i^n k_x^i + \frac{\beta^2 \rho_0 \hbar}{p_0}\sum_j^nk_z^j\left(\alpha - \sum_i^j \Delta \alpha_i\right)\\
-\frac{\hbar}{\rho_0}\sum_j^nk_z^j
  \end{pmatrix}. \label{full_shift_photon}
\end{align}
\end{widetext}

\section{\label{app:power_spectrum}Power Spectral Density}
The noise \(n(t)\) on the GW signal \(s(t)\) is measured in addition to the strain \(h(t)\), such that 
\begin{align}
s(t) &= h(t) + n(t).\nonumber
\end{align}
The one-sided power spectral density (PSD) of the noise \(S_n(f)\) is then given as
\begin{align}
S_n(f) &= \frac{2|\tilde{n}(f)|^2}{T}, \nonumber
\end{align}
where \(\tilde{n}(f) = \text{FT}[n(t)](f)\) is the Fourier transform of the noise time series and \(T\) the duration over which the noise is recorded \cite{Kaiser2021,Moore2015}. Since the gravitational strain is measured as time deviation \(\Delta T\) in the experiment, the resulting timing deviation has to be converted into effective noise amplitude. For a large number of photon emission, the time-of-flight signal due to synchrotron radiation consists of a predictable effective quadratic slow-down \(q_\text{eff}(t)\), see Eq.\ \ref{time_deviation_synch_rad}, and a small jitter \(n_T(t)\), such that the noise part of the signal consists of
\begin{align}
\Delta T_1 &=q_\text{eff}(t) + n_T(t).\nonumber
\end{align}
By taking the difference with the predicted quadratic slow-down \(\Delta T_2 = q(t)\) and performing the Fourier transform, in the case of sufficiently many emission events, the frequency components due to the timing of the emitted photons from their mean values remain in the PSD only. It is approximated, that no momentum deviation occurs, but all photons are emitted with the mean momentum. By inverting Eq.\ \ref{suvrat_signal}, the noise on the timing deviation \(n_T(t)\) can be related to noise in the GW strain \(n(t)\) via 
\begin{align}
\frac{1}{1-\frac{v_0^2}{2 c^2}}\frac{\text{d}}{\text{dt}}\left[q_\text{eff}(t) + n_T(t) - q(t) \right] &\approx n(t), \label{both_noises_relation}
\end{align}
which leads to the PSD of the GW strain noise
\begin{align}
S_n(f) &   \approx \frac{2}{T}\left|i 2 \pi f \frac{\text{FT}[(q_\text{eff}(t) + n_T(t) - q(t)) ](f)}{(1-\frac{v_0^2}{2 c^2})}\right|^2 \; 
\label{PSD_gw_strain}
\end{align}
where the additional factor arises due to the Fourier transform of the time derivative. Analytical expressions for this are derived by the use of a toy model, which will be discussed in the following.

\subsection{\label{appsup:toy_model_definition} Definition of a Toy Model}
It is assumed that a single photon emitted at time \(t_j\), leads to an acceleration \(a_j\) of a single particle via \(a(t) = a_j \delta(t-t_j)\). Thus, the particle accumulates a random momentum deviation over time and after the time interval \(\Delta t\) has passed, it is ejected and a new one inserted. Within the \(n\)-th \(\Delta t\)-interval \(k_n\) photons are emitted. By integrating this acceleration twice, and normalising it to \(1\) at the end of each time interval \(\Delta t\), the timing deviation for time \(t\) of the particle after emitting \(\sum_n k_n\) photons, is given as 
\begin{align}
\Delta T_1 &= \frac{2}{\Delta t}\sum_n^N \frac{1}{k_n}\sum_j^{k_n}(t- (n-1) \Delta t - t_j)\times  \nonumber \\
&\Theta(t- (n-1) \Delta t - t_j) \Theta(n \Delta t - t),\label{model_1}
\end{align}
where an additional sum over \(n\) has been introduced that accounts for dumping the particle every \(\Delta t\) and injecting a new one without initial timing deviation. In the following the number of emitted photons during each time interval \(\Delta t\) is approximated to be the same, such that \(k_n \approx k \; \forall n\). This model results in an effective quadratic dependency of the time delay \(\Delta T\) on the time \(t\). The average quadratic time delay is given by another model
\begin{align}
\Delta T_2 =\frac{1}{\Delta t^2} \sum_n^N & \left(t- (n-1) \Delta t\right)^2 \Theta(t - (n-1) \Delta t)\times \nonumber \\
&\Theta(n \Delta t - t),
\label{model_2} 
\end{align}
which is also normalised to \(1\) at the end of each time interval \(\Delta t\). In the following this normalisation is kept, but at the end of this section, the global factor relating these models to the noise expected in the storage ring is introduced.
\subsection{\label{appsup:toy_model_fourier_transform} Fourier Transform}
Using the Fourier transform of a function \(F(t)\)
\begin{align}
\tilde{F}(\omega) &= \frac{1}{\sqrt{2\pi}}\int_{-\infty}^{\infty} \text{d}t \; e^{- i \omega t} F(t),
\end{align}
Eq.\ \ref{model_1} is transformed into
\begin{align}
\tilde{q}_\text{eff}(f) + \tilde{n}_T(f) &\sim \frac{2}{k \Delta t }\sum_n^N \sum_j^k \left[ \frac{e^{-i \omega n \Delta t}}{\sqrt{2\pi}\omega^2} - \frac{e^{-i \omega (t_j + (n-1) \Delta t)}}{\sqrt{2\pi}\omega^2} \right. \nonumber \\
&\left.- \frac{i e^{-i \omega n \Delta t }t_j}{\sqrt{2\pi}\omega} +\frac{i e^{i \omega n \Delta t} \Delta t}{\sqrt{2\pi} \omega}\right], \label{fourier_transform_1}
\end{align}
which corresponds to the unitless part of the time delay obtained from individual photon emissions. If each photon is emitted on average after the duration \(\Delta t/k\), then the \(j\)th photon is emitted at \(t^0_j = j(\Delta t/k)\), such that the sum, together with the factor \(1/k\) in the limit \(k\rightarrow \infty\) turn into an integral, acting on Eq.\ \ref{fourier_transform_1}. This leads to 
\begin{align}
  &\lim_{k \rightarrow \infty} \left(\tilde{q}_\text{eff}(f) + \tilde{n}_T(f) \right) \sim \frac{1}{\Delta t^2}\sum_n^N\left[ \frac{-i e^{i n \omega \Delta t}\sqrt{\frac{2}{\pi}}}{\omega ^3} \right. \nonumber \\
  & \left. + \frac{i e^{-i\omega(n-1)\Delta t}}{\sqrt{2\pi}\omega^3}+\frac{\Delta t e^{-i \omega n \Delta t}\sqrt{\frac{2}{\pi}}}{\omega^2} + \frac{i \Delta t^2 e^{-i \omega n \Delta t}\sqrt{\frac{2}{\pi}}}{2\omega}\right] \label{fourier_transform_2},
\end{align}
which is exactly the Fourier transform of Eq.\ \ref{model_2} and which correponds to the unitless part of the model with the average time delay. Next, each photon is assumed to be emitted with a small timing deviation \(\varepsilon_j\) from the expected value, such that \(t_j =t^0_j + \varepsilon_j\), leading to some additional first-order terms in Eq.\ \ref{fourier_transform_1}. Therefore, for sufficiently many photon emission events
\begin{align} 
  \tilde{q}_\text{eff}(f) + \tilde{n}_T(f) - \tilde{q}(f) &\sim \frac{i}{k\Delta t} \left[\sum_n^N\sqrt{\frac{2}{\pi}}\frac{e^{-i \omega n \Delta t}}{\omega}\right] \times \nonumber \\
  &\sum_j^k \varepsilon_j \left(e^{i \omega (\Delta t - t_j^0)} - 1 \right) \label{noise_fourier_transform}
\end{align}
is the unitless part of the Fourier transform of the arrival time noise after subtraction of the expected quadratic result. This noise can be related via Eq.\ \ref{PSD_gw_strain} to the noise on the GW strain.

\begin{figure}[h]
  \includegraphics[width = 0.5\textwidth]{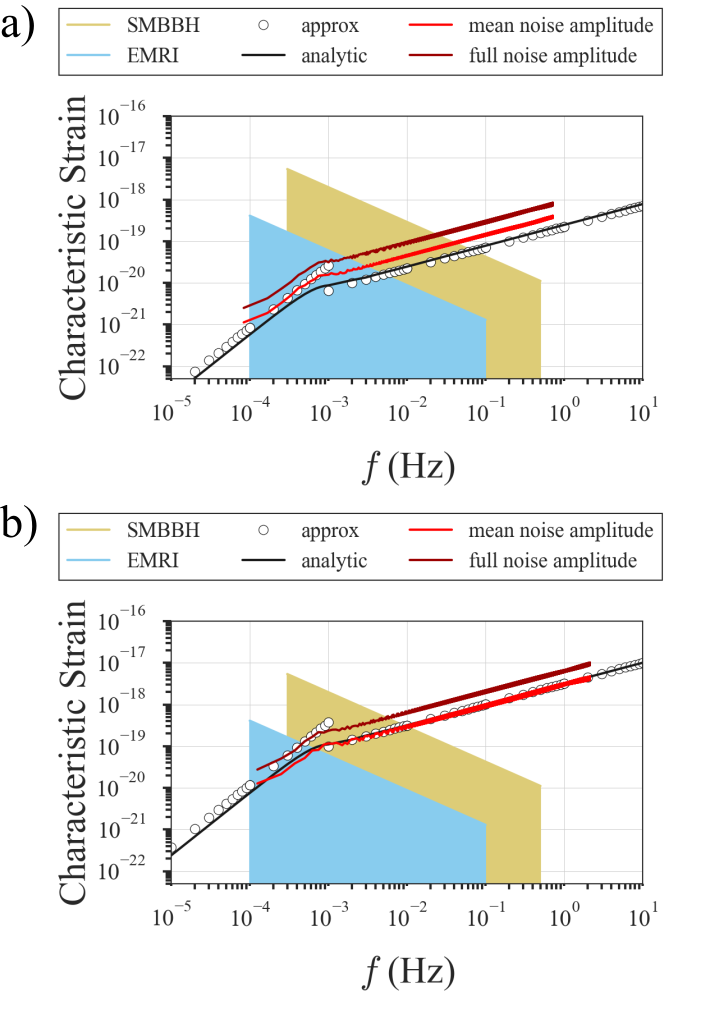}
  \caption{\label{fig:noise_sensitivity}Comparison of numerical results for the noise amplitude of \(\text{U}_2^{+/-}\) in a model storage ring with parameters stated in Tab.\ \ref{tab:table1}, where a) \(\beta=0.32\) and b) \(\beta = 0.95\). The characteristic strain of SMBBH and EMRI type sources is indicated, as well as the numerical results for noise amplitude of momentum and emission time sampling ("full"), and for only emission time sampling with mean momentum ("mean") in dark and light red. The analytical result for the PSD is shown as black line and an approximation to the analytical result as white circles.}
\end{figure}

\subsection{\label{appsup:toy_model_power_spectrum} Power Spectral Density}
By multiplying Eq.\ \ref{noise_fourier_transform} with the factor \(\frac{-i \omega}{1-\frac{v_0^2}{2c^2}}\) and then taking the absolute square, the normalised power spectral density of the GW strain noise is obtained. The exponential functions in the square brackets in Eq.\ \ref{noise_fourier_transform} are highly oscillating for frequencies around and above \(1/\Delta t\), such that the average value is taken as
\begin{align}
  \left|\sum_n^N e^{-i \omega n \Delta t}\right|^2 & \approx N \label{approx_exp_func}.
\end{align}
With this, the normalized part of the power spectral density of the fluctuations is given as
\begin{align}
  S_n(\omega) & \sim \frac{2}{T}\left(\frac{1}{1-\frac{v_0^2}{2c^2}}\right)^2\frac{N}{\pi k^2 \Delta t^2} \left(\left|\sum_j^k \varepsilon_j e^{-i \omega t_j^0} \right|^2 \right.  + \left|\sum_j^k \varepsilon_j \right|^2 \nonumber \\
  & \left.- \left( \sum_j^k \varepsilon_j\right) \sum_j^k 2\varepsilon_j \cos{(\Delta t - t_j^0)\omega}\right). \label{power_spectral_density_before_approximation}
\end{align}
In the form of Eq.\ \ref{power_spectral_density_before_approximation}, numerical values can only be produced, when the emission time is sampled by the Poisson distribution and a timing deviation \(\varepsilon_k\) for each emitted photon is obtained. Evaluating Eq.\ \ref{power_spectral_density_before_approximation} for \(100\) different sets of emission time samples, the power spectral density is obtained. It is turned into the correct units by a factor \(B^2\), where \(B\) is the omitted factor in Eq.\ \ref{noise_fourier_transform}. It originates from the average radiated power due to the synchrotron radiation (Eq.\ \ref{time_deviation_synch_rad}), turning Eqs.\ \ref{model_1} and \ref{model_2} into the time delay of a particle circulating in a storage ring after emission of individual photons, and the time delay due to the emission of the average power, respectively. Furthermore, the factor takes into account the net projection of photon emission in the longitudinal direction, \(S_0\) in Eq.\ \ref{net_factor_longitudinal}, as well as the factor that relates the characteristic noise \(n(t)\) to the time delay of the particle Eq. \ref{both_noises_relation}. It is given by
\begin{align}
  B &= \left(1-\frac{v_0^2}{2 c^2}\right)\frac{\langle P_\text{eff} \rangle}{2 p_0 c \gamma^2} S_0 \Delta t^2.
\end{align} 
In addition, the time interval over which the noise is considered is related to the time interval between injection, dumping and reinjection via \(T = N \Delta t\), and an additional factor \(2\pi\) is required to relate the analytical PSD to the numerical result in the following. Altogether the final result of the power spectral density is given by
\begin{align}
  S_n(f) & = \frac{4\Delta t}{k^2}S_0^2\left(\frac{\langle P_\text{eff} \rangle}{2 p_0 c \gamma^2}\right)^2 \left(\left|\sum_j^k \varepsilon_j e^{-i 2\pi f t_j^0} \right|^2 \right.  + \left|\sum_j^k \varepsilon_j \right|^2 \nonumber \\
  & \left.- \left( \sum_j^k \varepsilon_j\right) \sum_j^k 2\varepsilon_j \cos{(\Delta t - t_j^0)2 \pi f}\right). \label{power_spectral_density_before_approximation_correct_units}
\end{align}
The sky and polarization averaged sensitivity \(\mathfrak{R}\) of the detector, is derived using the formalism stated in \cite{Kaiser2021, Moore2015} and the antenna pattern for the "plus" polarization \({F_+ = {\sin^2{\theta}\cos{2\psi}}}\) of a storage ring, where \((\theta, \phi)\) denote the spherical polar angles and \(\psi\) the polarization angle \cite{Rao2020}. With
\begin{align}
  \mathfrak{R} &= \int_0^{2\pi}\frac{\text{d}\psi}{2\pi} \int_0^{2\pi} \frac{\text{d}\phi}{2\pi} \int_0^\pi \text{d}\theta \frac{\sin{\theta}}{2} F_+^2 
\end{align}
we obtain \(\mathfrak{R}=\frac{4}{15}\), which quantifies the instrument's response to an incident GW, required for computing the noise amplitude, which is defined via \(h_n(f) = \sqrt{f \frac{S_n(f)}{\mathfrak{R}}}\) \cite{Kaiser2021}. Using Eq.\ \ref{power_spectral_density_before_approximation_correct_units} the black curve in Fig.\ \ref{fig:noise_sensitivity} a) and b) as the numerical result of the average arrival time noise of \(100\) particles is shown for \(U_2^{+/-}\). In the figure the numerical result for the full photon shot noise is shown in darker color and in lighter color if all photons are emitted with the mean momentum, and only the emission time is statistically sampled -- leading to lower noise in general. For the slower particle with \(\beta=0.32\), shown in a), the analytical result of Eq.\ \ref{power_spectral_density_before_approximation_correct_units} lies below the numerical result of the noise amplitude, expected from emission with mean photon momenta. The reason is that the assumption of many photon emission events, leading to the cancellation of some terms in Eq.\ \ref{noise_fourier_transform} is not valid, as only very few photons are emitted during the circulating particles. Thus, the analytical result under estimates the noise for very little synchrotron emission. For much faster particles (\(\beta = 0.95\)), the analytical results captures the noise of photoemission with mean momenta very well, as it coincides with the numerical result, see Fig.\ \ref{fig:noise_sensitivity} b). For both particle velocities, the full noise amplitude, including momentum sampling, lies above the noise amplitude taking only mean momentum into account, as is expected.

\subsection{\label{appsup:toy_model_approximation} Approximation Formulas}

Eq.\ \ref{power_spectral_density_before_approximation_correct_units} is approximated for \(f\ll 1/\Delta t\) and \(f\gg 1/\Delta t\), leading to
\begin{align}
  S_n(f) &\approx\frac{4\Delta t}{k^2}S_0^2\left(\frac{\langle P_\text{eff} \rangle}{2 p_0 c \gamma^2}\right)^2 \times \nonumber \\
  &\begin{cases}
    \begin{tabular}{@{}c@{}}
    \(\left(2\pi f \Delta t \right)^2 \left\{k \langle \varepsilon \rangle \sum_j^k \varepsilon_j \left(1-j/k\right)^2 \right.\)
    \\
    \(\left.+ \left| \sum_j^k \varepsilon_j \left(1-j/k\right) \right|^2 \right\} \)
    \end{tabular} & f < 1/\Delta t \\
    \left(\left(\sum_j^k \varepsilon_j\right) - \langle \varepsilon \rangle\right)^2 & f > 1/\Delta t.
  \end{cases}\label{psd_first_approximation}
\end{align}
where \(\sum_j^k \varepsilon_j \approx k \langle \varepsilon\rangle\). This function indicates the two regimes shown as the approximation to the noise amplitude in Fig.\ \ref{fig:noise_sensitivity}, denoted as \emph{approx.} in white circles and shows a proportionality \(\sqrt{\frac{f S_n(f)}{\mathfrak{R}}}\sim f^{3/2}\) for \(f < 1/\Delta t\) and \(\sim f^{1/2}\) for \(f > 1/\Delta t\), clearly in line with the numerical computation. Next \(\sum_j^k \varepsilon_j \frac{j}{k} \approx
\sum_j^{k/2} \varepsilon_j  = \langle \varepsilon\rangle k/2\) is crudely estimated. This leads to 
\begin{align}
  S_n(f) &\approx \frac{4\Delta t}{k^2}S_0^2\left(\frac{\langle P_\text{eff} \rangle}{2 p_0 c \gamma^2}\right)^2 k^2 \langle \varepsilon \rangle^2 \times \nonumber \\
  &\begin{cases}
    \left(2\pi f \Delta t\right)^2 /4 & f < 1/\Delta t \\
    1  & f > 1/\Delta t
  \end{cases}\label{psd_second_approximation}
\end{align}
as final result for the approximate power spectral density of the synchrotron radiation.
\subsection{\label{numerical_PSD} Power Spectral Density from Arrival Time Simulation}
For simulating the arrival time noise, \(n\) \(\text{U}^{+/-}\) or \(\text{U}_2^{+/-}\) ions are tracked in the model storage ring, and their accumulated time deviation is reset after \(\Delta t = 10^3\) s, corresponding to dumping and reinsertion. The resulting time deviation is transformed via Eq.\ \ref{relation_signal_strength_main} to that of a ring with fixed radius and the quadratic offset is subtracted for each of the time intervals, leading to a measurement time series \(\Delta T_i(t)\) for the \(i\)th ion, which corresponds to \(q_\text{eff}(t) + n_T(t) - q(t)\) in the analytic model. The noise amplitude is obtained via Eq.\ \ref{both_noises_relation}, from which the numerical derivative of the time series is computed. Then a fast Fourier transform is applied to the result, where \(\Delta \tilde{t}\) is the resolution of the time series and \(\langle \dots\rangle\) is the average over \(i\). It leads to the numerical power spectral density given by
\begin{widetext}
\begin{equation} 
S_n^\text{sim}(f_j) = \frac{2\Delta \tilde{t}}{N_k}\left\langle \left|\frac{1}{1-\frac{v_0^2}{2c^2}} \sum_{j=1}^{N_k-1} e^{-\frac{i2\pi f j}{N_k}}\frac{\Delta T(t_{j+1})-\Delta T(t_j)}{t_{j+1} - t_{j}} \right|^2 \right\rangle.
\end{equation}
\end{widetext}

\bibliographystyle{apsrev}

\end{document}